\documentclass[journal,twoside]{IEEEtran}
\usepackage{graphicx,cite,amssymb,amsmath,bm}
\usepackage{mathrsfs}
\usepackage{color,soul}
\usepackage{tabulary}
\usepackage{multirow,multicol}
\usepackage{cases}
\usepackage{stfloats}
\usepackage{url}
\usepackage{booktabs}
\usepackage{algpseudocode}
\usepackage{algorithm,algpseudocode}
\usepackage{algorithmicx}
\usepackage[normalem]{ulem}
\ifCLASSOPTIONcompsoc
    \usepackage[caption=false, font=normalsize, labelfont=sf, textfont=sf, subrefformat=parens, labelformat=parens]{subfig}
\else
\usepackage[caption=false, font=footnotesize, subrefformat=parens, labelformat=parens]{subfig}
\fi

\usepackage{amsmath,amssymb}
\usepackage{mathtools} 
\usepackage{xspace}
\newcommand{\Tableref}[1]{Table~\ref{#1}}

\newcommand{\eqrefs}[2]{equations~(\ref{#1})-(\ref{#2})}
\newcommand{\Secref}[1]{Section~\ref{#1}}
\newcommand{\Figref}[1]{Fig.~\ref{#1}}

\newcommand{\herm}{^\text{H}}

\newcommand{\trans}{^\text{T}}

\newcommand{\bx}{\mathbf{x}}
\newcommand{\bz}{\mathbf{z}}

\newcommand{\bs}{\mathbf{s}}

\newcommand{\bX}{\mathbf{X}}

\newcommand{\bg}{\mathbf{g}}

\newcommand{\bw}{\mathbf{w}}
\newcommand{\bY}{\mathbf{Y}}
\newcommand{\by}{\mathbf{y}}
\newcommand{\bI}{\mathbf{I}}

\newcommand{\bb}{\mathbf{b}}

\newcommand{\bv}{\mathbf{v}}
\newcommand{\bU}{\mathbf{U}}
\newcommand{\bu}{\mathbf{u}}

\newcommand{\bOmega}{\boldsymbol{\Omega}}
\newcommand{\bww}{\boldsymbol{\omega}}

\newcommand{\bvphi}{\boldsymbol{\varphi}}
\newcommand{\bpsi}{\boldsymbol{\psi}}

\newcommand{\bzero}{\boldsymbol{0}}

\newcommand{\CN}{\mathcal{CN}}
\newcommand{\tc}{\tau_{\textsc{c}}}

\newcommand{\boundellipse}[3]% center, x rad, y rad
{(#1) ellipse [x radius=#2,y radius=#3]
}

\DeclareMathOperator{\Tr}{Tr}

\DeclareMathOperator{\var}{\mathsf{Var}}

\newcommand{\EX}[1]{\mathsf{E}\left\{{#1}\right\}}

\newcommand{\varx}[1]{\var\left\{{#1}\right\}}

\newcommand{\cov}[1]{\mathsf{Cov}\left\{{#1}\right\}}

\newcommand{\C}{\mathbb{C}}

\newcommand{\Pu}{\rho_{\mathrm{u}}}

\newcommand{\Pd}{\rho_{\mathrm{d}}}
\newcommand{\Pdp}{\rho_{\mathrm{d,p}}}

\newcommand{\tauc}{\tau_\mathrm{c}}

\newcommand{\tauup}{\tau_\mathrm{u,p}}

\newcommand{\taudp}{\tau_\mathrm{d,p}}

\newcommand{\norm}[1]{{ \left\Vert #1 \right\Vert }}

\newcommand{\ECB}{^\textsf{ECB}}
\newcommand{\CBDT}{^\textsf{CB-DT}}
\newcommand{\CB}{^\textsf{CB}}
\newcommand{\NCB}{^\textsf{NCB}}
\newcommand{\iid}{\text{i.i.d.}}

\makeatletter
\def\@setsize#1#2#3#4{
    \@nomath#1
    \let\@currsize#1
    \baselineskip #2
    \baselineskip \baselinestretch\baselineskip
    \parskip \baselinestretch\parskip
    \setbox\strutbox \hbox{
        \vrule height.7\baselineskip
            depth.3\baselineskip
            width\z@}
    \skip\footins \baselinestretch\skip\footins
    \normalbaselineskip\baselineskip#3#4}
\makeatother

\makeatletter
\newcommand{\setstretch}[1]{
    \def\baselinestretch{#1}%
    \@currsize
    }
\makeatother



\def\BibTeX{{\rm B\kern-.05em{\sc i\kern-.025em b}\kern-.08em
    T\kern-.1667em\lower.7ex\hbox{E}\kern-.125emX}}

\newcommand{\myfrac}[3][0pt]{\genfrac{}{}{}{}{\raisebox{0pt}{$#2$}}{\raisebox{-#1}{$#3$}}}

\newcounter{eqcnt1}
\newcounter{eqcnt2}
\newcounter{eqcnt3}
\newcounter{eqcnt4}

\newcommand{\comment}[1]{}

\begin{document}
%\linenumbers
\begin{figure*}[t!]
\normalsize
This paper was submitted for publication in IEEE Transactions on Communications on August 28, 2020. It was finally accepted for publication on January 21, 2021.

\

\textcopyright~2021 IEEE.  Personal use of this material is permitted.  Permission from IEEE must be obtained for all other uses, in any current or future media, including reprinting/republishing this material for advertising or promotional purposes, creating new collective works, for resale or redistribution to servers or lists, or reuse of any copyrighted component of this work in other works.
\vspace{20cm}
\end{figure*}

\newpage

\title{Enhanced Normalized Conjugate Beamforming \\ for Cell-Free Massive MIMO}
\author{Giovanni~Interdonato,~\IEEEmembership{Member,~IEEE,} Hien~Quoc~Ngo,~\IEEEmembership{Senior Member,~IEEE} and~Erik~G.~Larsson,~\IEEEmembership{Fellow,~IEEE}%
\thanks{
        G.~Interdonato was with the Dept. of Electrical Engineering (ISY), Link\"{o}ping University, 581 83 Link\"{o}ping, Sweden, and is now with the Dept. of Electrical and Information Engineering, University of Cassino and Southern Lazio, 03043 Cassino, Italy (giovanni.interdonato@unicas.it)
        }
\thanks{
        H.~Q.\ Ngo is with the Institute of Electronics, Communications and Information Technology (ECIT), Queen's University Belfast, Belfast, BT3 9DT, U.K. (hien.ngo@qub.ac.uk).}
\thanks{
        E.~G. Larsson is with the Dept. of Electrical Engineering (ISY), Link\"{o}ping University, 581 83 Link\"{o}ping, Sweden (erik.g.larsson@liu.se).}
\thanks{The work of G.~Interdonato and E.~G.~Larsson was supported in part by the Swedish Research Council (VR) and ELLIIT. The work of H.~Q. Ngo was supported by the U.K. Research and Innovation Future Leaders Fellowships under Grant MR/S017666/1.}
\thanks{Part of this work was presented at the 2016 IEEE 21st International Workshop on Computer Aided Modelling and Design of Communication Links and Networks (CAMAD)~\cite{Interdonato2016a}.}
}

\markboth{IEEE Transactions on Communications,~Vol.~XX, No.~X, XXXX}%
{Interdonato \MakeLowercase{\textit{et al.}}: Enhanced Normalized Conjugate Beamforming for Cell-Free Massive MIMO}

\maketitle

\begin{abstract}
In cell-free massive multiple-input multiple-output (MIMO) the fluctuations of the channel gain from the access points to a user are large due to the distributed topology of the system. Because of these fluctuations, data decoding schemes that treat the channel as deterministic perform inefficiently.
A way to reduce the channel fluctuations is to design a precoding scheme that equalizes the effective channel gain seen by the users. Conjugate beamforming (CB) poorly contributes to harden the effective channel at the users. In this work, we propose a variant of CB dubbed \textit{enhanced normalized} CB (ECB), in that the precoding vector consists of the conjugate of the channel estimate normalized by its \textit{squared} norm. For this scheme, we derive an exact closed-form expression for an achievable downlink spectral efficiency (SE), accounting for channel estimation errors, pilot reuse and user's lack of channel state information (CSI), assuming independent Rayleigh fading channels. We also devise an optimal max-min fairness power allocation based only on large-scale fading quantities. ECB greatly boosts the channel hardening enabling the users to reliably decode data relying only on statistical CSI. As the provided effective channel is nearly deterministic, acquiring CSI at the users does not yield a significant gain.
\end{abstract}


\section{Introduction} \label{sec:intro}
\IEEEPARstart{C}{ell-free} massive multiple-input multiple-output (MIMO)~\cite{Ngo2017b,Interdonato2019,Zhang2019b} is a practical and scalable embodiment of network MIMO, and promises unprecedented levels of SE by leveraging an extraordinary macro-diversity and an aggressive spatial multiplexing of the users.
The practicality and the scalability of such a system comes from decentralizing channel estimation and precoding/combining, enabled by operating in time division duplex (TDD). The distributed dense topology of cell-free massive MIMO enriches the macro-diversity and allows to implement a \textit{user-centric} network wherein every user is in the center of a tailored virtual cell surrounded by serving cooperating access points (APs).
On the other hand, since the APs are geographically spread out, they contribute quite differently to the effective fading channel gain seen at each user, which is thereby characterized by large fluctuations. Hence, the channel is far to be deterministic. This behavior does not occur in co-located massive MIMO where the channel is nearly deterministic instead, under most relevant operating conditions~\cite{redbook,massivemimobook,Heath2018}, a phenomenon known as \textit{channel hardening}~\cite{Hochwald2004,Martinez2016,Gunnarsson2020}.

The lower degree of channel hardening in cell-free massive MIMO compared to co-located massive MIMO was pointed out in~\cite{Interdonato2019b}, and also analytically demonstrated in~\cite{ZChen2018,Polegre2020,Polegre2020b} under different channel model assumptions. As the channel does not sufficiently harden, the lack of CSI at the user constitutes a significant limitation in the performance. In this regard, a scalable pilot-based downlink training scheme for cell-free massive MIMO was advocated in~\cite{Interdonato2019b} to let the users perform data decoding based on the acquired CSI rather than relying on the statistical~CSI.

Since deriving from the \textit{law of the large numbers}, the channel hardening property depends on the number of antennas in the system. In general, the more antennas the more the channel hardens. Importantly, channel hardening at the users does also depend on the adopted precoding scheme as the effective downlink channel gain is given by the inner product between the downlink channel vector and the precoding vector. Hence, the channel hardening can be artificially boosted by acting on the precoding scheme. In our preliminary work~\cite{Interdonato2016a}, we proposed a different flavour of the conventional conjugate beamforming (CB) dubbed \textit{normalized} CB (NCB) for cell-free massive MIMO systems with single-antenna APs. With NCB, the precoding factor (in this case a scalar) consists of the conjugate of the channel estimate normalized by its magnitude. This scheme enables a reduction of the uncertainty due to the user's lack of CSI knowledge which in turn improves the SE. Recently, the authors in~\cite{Polegre2020} have extended the analysis of~\cite{Interdonato2016a} to multi-antenna APs, providing an exact closed-form expression for an achievable downlink SE based on the popular \textit{hardening} bound~\cite{redbook,massivemimobook}. Another modified CB scheme for cell-free massive MIMO was proposed in~\cite{Attarifar2019}, where the global CSI knowledge at the APs is exploited to compensate for the channel fluctuations and focus the overall channel gain around a desired mean target.

\textbf{Contribution:} In this paper, we propose a variant of the NCB precoding scheme described in~\cite{Interdonato2016a,Polegre2020} dubbed enhanced NCB (ECB), where the vector of the channel estimates between a multi-antenna APs and a given user is normalized by its squared norm. We provide an exact closed-form expression for an achievable downlink SE by using the popular hardening bound. This expression accounts for channel estimation errors at the AP, pilot contamination due to pilot reuse, and lack of CSI at the user side. We assume independent Rayleigh fading channels which is the best fading scenario for a channel hardening perspective~\cite{Ngo2017a} and allows us to draw insightful conclusions by inspecting the elegant derived closed-form SE expression. Based on the latter, which depends only on the large-scale fading quantities, we devise an optimal max-min fairness (MMF) power allocation scheme re-adapting the convex optimization framework used in~\cite{Ngo2017b}. This policy demands for a centralized coordination but, importantly, does not depend on the small-scale fading realizations. Hence, it can be performed by a central processor where the channel statistics are reasonably assumed to be available. 
The novelty of this study consists of:
\begin{itemize}
\item We provide a comprehensive SE analysis for a normalized CB scheme where the normalization term in the precoding vector (which characterizes an AP-user pair) is the \emph{squared} norm of the respective channel estimate, rather than just the norm of it, as in~\cite{Interdonato2016a,Polegre2020}. Our choice guarantees a better channel gain equalization at the users, hence the term \textit{enhanced} CB. This analysis is novel in the context of cell-free massive MIMO. A similar precoder was proposed in~\cite{Sutton2019} for co-located massive MIMO systems, but its analysis does not consider power control and pilot contamination. 

\item We devise a ``local'' solution to greatly boost the channel hardening. A substantial difference between the proposed ECB and the modified CB in~\cite{Attarifar2019} is that the latter requires CSI exchange among the APs which  scales unfavorably as the number of users and APs in the system grows. Moreover, the ``local'' nature of ECB, namely that each AP only needs its own channel estimates to construct the precoding vectors, is preferable in applications where latency is a concern and/or in systems with constrained fronthaul network capacity where the additional overhead due to the CSI exchange cannot be afforded.
   
\item We give a rigorous formulation of the MMF power control optimization problem for both ECB and NCB. The latter extends the power control {{analysis}}~of~\cite{Interdonato2016a,Polegre2020}.

\item We derive an approximate closed-form expression of an achievable downlink SE for CB assuming downlink training and multi-antenna APs along with the corresponding formulation of the MMF power control optimization problem. These results  extend those in \cite{Interdonato2019b}, and serve  as an upper bound in the performance evaluation.

\end{itemize}

\textbf{Related work:} As introduced earlier, the only studies investigating the beamforming normalization are~\cite{Interdonato2016a,Sutton2019,Attarifar2019,Polegre2020}. However, cell-free massive MIMO has recently received great attention, and in general is a large research topic. 
The research conducted over the last few years aimed to analyze this concept and evaluate its performance in practical implementations. \cite{Borg2018,Bjornson2020,Riera2020b} focus on the scalability and decentralization aspects of cell-free massive MIMO, while~\cite{Atzeni2020,Gouda2020} propose an over-the-air signalling scheme to avoid the exchange of CSI for centralized baseband processing. The dynamic clustering approach enabling the user-centric network implementation is discussed in~\cite{Yuan2017c,Pan2018b,Buzzi2019c}. Optimal and heuristic downlink power allocation algorithms are devised in~\cite{Nayebi2017,Francis2019,Nikbakht2020} operating in centralized and distributed fashion, respectively. The effectiveness of minimum mean-square error (MMSE) combining with large-scale fading decoding (LSFD) and suboptimal AP selection policies has been established in~\cite{Nayebi2016,Bjornson2020a} and~\cite{Attarifar2020}, respectively. Finally, a significant effort has been spent to evaluate the performance of cell-free massive MIMO under realistic operating assumptions: finite fronthaul capacity~\cite{Parida2018,Bashar2019c}; hardware impairments~\cite{Zhang2018,Masoumi2020}; low-resolution analog-to-digital converters~\cite{yZhang2019,Hu2019}; and imperfect channel reciprocity~\cite{Palacios2020}.     

\section{System Model} \label{sec:sysmodel}
Let us consider a cell-free massive MIMO system with $M$ multi-antenna APs providing service to $K$ single-antenna users in the same time-frequency resources. Each AP is equipped with $N$ antennas, and it holds $MN \gg K$. A central processing unit (CPU) masters all the APs through a fronthaul network. It is responsible for data sharing, clock synchronization and centralized operation for resource allocation tasks.

The system operates in TDD mode and we assume block-fading model. The time-frequency resources are structured in coherence blocks wherein the channel is approximately static and frequency flat. The TDD coherence block is $\tauc$ samples long and is determined by the shortest user's coherence time and bandwidth in the system, as $\tauc = T_{\mathrm{c}} B_{\mathrm{c}}$. Conventionally, a coherence block accommodates three phases: $(i)$ uplink training, $(ii)$ uplink data transmission\footnote{The uplink data transmission phase is out of the scope of this work and thereby its analysis is herein omitted.}, and $(iii)$ downlink data transmission.

Let $\bg_{mk} \in \C^{N}$ denote\footnote{\Tableref{tab:notation} summarizes the most relevant notation.
} the channel response vector between user $k$ and multi-antenna AP $m$. We assume independent Rayleigh fading channels, thereby $\bg_{mk} \sim \CN(\bzero,\beta_{mk}\bI_N)$, where $\beta_{mk}$ is the large-scale fading coefficient capturing the path loss and the effects of correlated shadowing.

\begin{table*}[!t]
\centering
\normalsize
\renewcommand{\arraystretch}{1.1} 
\caption{List of Relevant Notations}
\begin{tabular}{|c|c|c|c|c|}
\hline 
 $M$ & 	number of APs & $\tauc$  & length of the coherence block  \\ \hline
 $N$ &  number of AP's antennas  & $\tauup$ & uplink training length \\ \hline
 $K$ &  number of users & $\taudp$ & downlink training length \\ \hline
 $\bg_{mk}$ & channel response vector between user $k$ and AP $m$ & $\Pu$ & normalized SNR of the uplink pilot symbol \\ \hline
 $\beta_{mk}$ & mean-square of any element of $\bg_{mk}$&  $\Pdp$ & normalized SNR of the downlink pilot symbol \\ \hline
$\hat{\bg}_{mk}$& MMSE estimate of $\bg_{mk}$ & $\Pd$ & normalized SNR of the downlink data symbol \\ \hline
$\gamma_{mk}$ & mean-square of any element of $\hat{\bg}_{mk}$ & $\bvphi_k$ & uplink pilot sequence related to user $k$ \\ \hline
$\tilde{\bg}_{mk}$& channel estimation error, $\bg_{mk}-\hat{\bg}_{mk}$  & $\bpsi_k$ & downlink pilot sequence related to user $k$ \\ \hline
$a_{kk}$ & effective downlink channel at user $k$ & $\bx_m$ & data signal transmitted by AP $m$ \\ \hline
$\hat{a}_{kk}$ & MMSE estimate of $a_{kk}$ & $\bw_{mk}$& precoding vector used by AP $m$ to user $k$ \\ \hline
$\tilde{a}_{kk}$& downlink channel estimation error, $a_{kk}-\hat{a}_{kk}$ & $\eta_{mk}$ & downlink power control coefficient \\ \hline 
\end{tabular}
\label{tab:notation}
\end{table*}

\subsection{Uplink Training}
Uplink training takes place via pilot transmission. The length of the pilot, $\tauup$ samples, determines the training duration as well as the number of mutually orthogonal pilots. Ideally, every user should use an orthogonal pilot sequence to prevent interference from pilot contamination in the channel estimates. However, the share of coherence block reserved to the training is limited and pilot reuse is unavoidable. Let $\sqrt{\tauup} \bvphi_k \in \C^{\tauup}$ be the pilot sequence sent by user $k$, $\norm{\bvphi_k}=1$. We assume that the pilots of any pair of users can be either identical or orthogonal, i.e., for any $k \neq j$ it~holds  
\begin{equation} \label{eq:pilot-design}
\bvphi_k\trans \bvphi^\ast_j = 
\begin{cases}
1, &\quad \text{if } \bvphi_k = \bvphi_j, \\
0, &\quad \text{otherwise.}
\end{cases}
\end{equation}
The overall pilot signal received at AP $m$ is given by
\begin{align}
\bY_{\mathrm{p},m} = \sqrt{\tauup \Pu}~\sum\nolimits_{k=1}^K \bg_{mk} \bvphi\trans_k + \bOmega_{\mathrm{p},m} \in \C^{N \times \tauup},
\end{align}
where $\Pu$ is the normalized signal-to-noise ratio (SNR) of the uplink pilot symbol, and $\bOmega_{\mathrm{p},m}$ is a matrix of additive noise whose elements are independent and identically distributed (\iid) $\CN(0,1)$. AP $m$ de-spreads the pilot signal by using the pilot sequences as
\begin{align}
\by_{\mathrm{p},mk} &\triangleq \! \bY_{\mathrm{p},m}\bvphi^\ast_k \nonumber \\
&= \! \sqrt{\tauup \Pu} \bg_{mk} \!+\! \sqrt{\tauup \Pu} \sum\limits_{j \neq k}^K \bg_{mj} \bvphi\trans_j \bvphi^\ast_k \!+\! \bww_{\mathrm{p},mk}
\end{align}
$\in \C^N$, where $\bww_{\mathrm{p},mk}\!=\!\bOmega_{\mathrm{p},m} \bvphi^\ast_k\!\sim\!\CN(\bzero, \bI_N)$. Due to the pilot sequence design in~\eqref{eq:pilot-design}, $\by_{\mathrm{p},mk}$ constitutes a sufficient statistic. Provided that $\{ \beta_{mk} \}$ are known a priori at the AP, linear MMSE channel estimation can be performed, and it is optimal---in the sense that it minimizes the mean square error---due to the Gaussian distribution of the channels. 
Channel estimation is carried out locally at each AP and the MMSE estimate of $\bg_{mk}$ is given~by
\begin{align}
\hat{\bg}_{mk} &= \EX{\bg_{mk}\by\herm_{\mathrm{p},mk}}\left( \EX{\by_{\mathrm{p},mk} \by\herm_{\mathrm{p},mk}} \right)^{-1} \by_{\mathrm{p},mk} \nonumber \\
&= c_{mk} \by_{\mathrm{p},mk},
\end{align}
where 
\begin{equation}
c_{mk} \triangleq \frac{\sqrt{\tauup\Pu} \beta_{mk}}{\tauup\Pu \sum\nolimits^K_{j=1} \beta_{mj} |\bvphi_k\herm \bvphi_j|^2 + 1}.
\end{equation}
The MMSE channel estimate is distributed as $$\hat{\bg}_{mk}\sim \CN(\bzero,\gamma_{mk}\bI_N)$$ where
\begin{equation}\label{eq:gamma}
\gamma_{mk} = \sqrt{\tauup\Pu} c_{mk} \beta_{mk}.
\end{equation}
In the channel estimate, the interference from pilot contamination is captured by the terms that are proportional to $\bvphi_k\herm \bvphi_j$ and for which this inner product gives 1. Note that, if user $k$ and user $j$ share the same pilot, then the respective channel estimates as well as their mean-squares are proportional to each other:
\begin{align} 
\hat{\bg}_{mk} = \dfrac{\beta_{mk}}{\beta_{mj}} \hat{\bg}_{mj}, \label{eq:correlated-estimates} \\
\gamma_{mk} = \dfrac{\beta^2_{mk}}{\beta^2_{mj}} \gamma_{mj} \label{eq:correlated-gammas}.
\end{align}
Finally, let us define the channel estimation error $$\tilde{\bg}_{mk} = \bg_{mk} - \hat{\bg}_{mk},$$ which is distributed as $$\tilde{\bg}_{mk} \sim \CN(\bzero,(\beta_{mk}-\gamma_{mk})\bI_N),$$ and independent of $\hat{\bg}_{mk}$. 

\subsection{Downlink Data Transmission}
By leveraging the channel reciprocity deriving from the TDD operation, the channel estimates obtained in the uplink are then employed to construct the precoding vectors.
Let $\bw_{mk} \in \C^N$ be the precoding vector used by AP $m$ in the service of user $k$. We assume that $\bw_{mk}$ solely depends on local channel estimates. Hence, there is no CSI exchange among the APs over the fronthaul network. 
The data signal that AP $m$ transmits to the users is 
\begin{equation} \label{eq:data-transmission}
\bx_m = \sqrt{\Pd} \sum\nolimits^K_{k=1} \sqrt{\eta_{mk}} \bw_{mk} q_k,
\end{equation}
where $q_k$ is the data symbol intended for user $k$, $\EX{|q_k|^2} = 1$ and $\EX{q_k q^\ast_j} = 0$ for $k \neq j$. $\Pd$ is the normalized SNR of the downlink data symbol, and $\{\eta_{mk}\}$ are power control coefficients satisfying the per-AP power constraint
\begin{equation}\label{eq:power-constraint}
\EX{\norm{\bx_m}^2} \leq \Pd,~m=1,\ldots,M.
\end{equation}
By setting $\eta_{mk} = 0$, AP $m$ does not participate in the service of user $k$. Hence, these coefficients are also useful to set up clustering policies aiming to preserve the scalability of the system.
 
\noindent The signal received at user $k$ resulting from the joint coherent transmission by multiple APs is
\begin{align} \label{eq:data-symbol}
r_k &= \sum\limits^M_{m=1} \bg\trans_{mk} \bx_m + \omega_k \nonumber \\
&= \sqrt{\Pd} \sum^M_{m=1} \sqrt{\eta_{mk}} \bg\trans_{mk} \bw_{mk} q_k \nonumber \\ 
&\quad+ \sqrt{\Pd} \sum^K_{j \neq k} \sum^M_{m=1} \sqrt{\eta_{mj}} \bg\trans_{mk} \bw_{mj} q_j + \omega_k,  
\end{align}
where $\omega_k\sim\CN(0,1)$ is additive noise. 

\begin{figure*}[!b]
\normalsize
\setcounter{eqcnt1}{\value{equation}}
\setcounter{equation}{17}
\hrulefill
\begin{equation}
\label{eq:SINR:CB}
\mathsf{SINR}\CB_k = \frac{\Pd N^2 \left( \sum\limits_{m=1}^M \sqrt{\eta_{mk}} \gamma_{mk} \right)^2}{\Pd N \sum\limits_{j = 1}^K \varsigma_{kj} + \Pd N^2 \sum\limits_{j \neq k}^K \left(\sum\limits_{m=1}^M \sqrt{\eta_{mj}} \gamma_{mj} \dfrac{\beta_{mk}}{\beta_{mj}} \right)^2\!\left|\bvphi_k\herm\bvphi_j\right|^2 + 1},
\end{equation}
\hrule
\vspace*{4pt}
\setcounter{equation}{20}
\begin{equation} \label{eq:SINR:NCB}
\mathsf{SINR}\NCB_k = \frac{\Pd \alpha^2 \left(\sum\limits_{m=1}^M \sqrt{\eta_{mk} \gamma_{mk}}\right)^2}{\Pd (N-1-\alpha^2) \sum\limits_{m=1}^M \eta_{mk} \gamma_{mk}+\Pd \sum\limits_{j=1}^K \sum\limits_{m=1}^M \eta_{mj} \beta_{mk}+\Pd \sum\limits_{j \neq k}^K \Upsilon_{kj}~|\bvphi\herm_k \bvphi_j|^2+1},
\end{equation}
where
\begin{align}
\Upsilon_{kj} &\triangleq (N-1) \sum\limits_{m=1}^M \eta_{mj} \gamma_{mj} \frac{\beta^2_{mk}}{\beta^2_{mj}} + \alpha^2 \sum\limits_{m=1}^M \sum\limits_{n \neq m}^M \sqrt{\eta_{mj}\eta_{nj}\gamma_{mj}\gamma_{nj}}\frac{\beta_{mk}\beta_{nk}}{\beta_{mj}\beta_{nj}}, \label{eq:NCB:upsilon}\\
\alpha &\triangleq \frac{\Gamma{(N+1/2)}}{\Gamma{(N)}}. \label{eq:alpha}
\end{align}
\setcounter{equation}{\value{eqcnt1}}
\end{figure*}

\section{Performance Analysis - No CSI at the User} \label{sec:performance-analysis}
When evaluating the capacity that this system can achieve, we must bear in mind the lack of CSI at the user, which employs the channel statistics to perform data decoding.
An achievable downlink spectral efficiency for user $k$ can be obtained by using the popular \textit{hardening} bound~\cite{redbook,massivemimobook}. We rewrite~\eqref{eq:data-symbol} as
\begin{align} \label{eq:data-symbol-capacity}
r_k = \mathsf{DS}_k q_k + \mathsf{BU}_k q_k + \sum\nolimits_{j \neq k}^K \mathsf{UI}_{kj} q_j + \omega_k,
\end{align}
where 
\begin{align}
\mathsf{DS}_k &= \sqrt{\Pd}\sum\nolimits^M_{m=1} \sqrt{\eta_{mk}}~\EX{\bg\trans_{mk} \bw_{mk}}, \label{eq:DS} \\
\mathsf{BU}_k &= \sqrt{\Pd} \sum\nolimits^M_{m=1} \sqrt{\eta_{mk}}~\bg\trans_{mk} \bw_{mk} \nonumber \\ 
&\quad - \sqrt{\Pd} \sum\nolimits^M_{m=1} \sqrt{\eta_{mk}}~\EX{\bg\trans_{mk} \bw_{mk}}, \label{eq:BU} \\
\mathsf{UI}_{kj} &= \sqrt{\Pd} \sum\nolimits^M_{m=1} \sqrt{\eta_{mj}} \bg\trans_{mk} \bw_{mj}, \label{eq:UI} 
\end{align}
emphasizing that user $k$ can detect $q_k$ by exploiting only the knowledge of the channel statistics, that is the knowledge of $\EX{\bg\trans_{mk} \bw_{mk}}$. Hence, $\mathsf{DS}_k$ represents the desired signal for user $k$, $\mathsf{BU}_k$ (beamforming gain uncertainty) is a self-interference contribution capturing user's uncertainty of the instantaneous channel gain, while $\mathsf{UI}_{kj}$ describes the inter-user interference. 
The term $\mathsf{BU}_k$   constitutes a measure of the channel hardening degree at user $k$.
Specifically, $\mathsf{BU}_k$ quantifies how much the instantaneous effective downlink channel deviates from its mean. 
The smaller $\mathsf{BU}_k$ is, the more hardening the channel offers.  
By treating the sum of the last three terms in~\eqref{eq:data-symbol-capacity} as uncorrelated effective noise, a lower bound on the downlink capacity is given~by
\begin{equation} \label{eq:SE}
\mathsf{SE}_k = \xi \left(1 - \frac{\tauup}{\tauc} \right) \log_2 (1+\mathsf{SINR}_k),
\end{equation}
where $0<\xi<1$ is the share of the coherence block reserved to the downlink, the pre-log factor $1 - \tauup/\tauc$ accounts for the pilot overhead, and the signal-to-interference-plus-noise ratio (SINR) at user $k$ is
\begin{equation} \label{eq:SINR}
\mathsf{SINR}_k = \frac{|\mathsf{DS}_k|^2}{\EX{|\mathsf{BU}_k|^2} + \sum\nolimits_{k \neq j}^K \EX{|\mathsf{UI}_{kj}|^2}  + 1}.
\end{equation}
Expression~\eqref{eq:SINR} is valid for any precoding scheme and channel model. We next report the closed-form expression of $\mathsf{SINR}_k$ assuming multi-antenna APs and independent Rayleigh fading channels for both CB~\cite{Ngo2018a}, NCB~\cite{Polegre2020} and the proposed ECB.

\subsection{Conjugate Beamforming} \label{sec:CB}
CB consists in setting $\bw\CB_{mk} = \hat{\bg}^\ast_{mk}$. By inserting this into~\eqref{eq:SINR} and computing the corresponding expectations, the effective SINR per user is given by~\cite{Ngo2018a} and~\eqref{eq:SINR:CB}, shown at the bottom of this page, where \addtocounter{equation}{1}
\begin{equation} \label{eq:variance:CB}
\varsigma_{kj} \triangleq \sum\nolimits_{m=1}^M \eta_{mj} \beta_{mk} \gamma_{mj}.
\end{equation}
Importantly, $\Pd N \varsigma_{kk}$ is the power of the beamforming gain uncertainty which equals the variance of the effective downlink channel, $\sum_{m=1}^M \sqrt{\Pd \eta_{mk}} \bg\trans_{mk} \hat{\bg}^\ast_{mk}$. Finally, by inserting $\bw_{mk} = \bw\CB_{mk}$ into~\eqref{eq:data-transmission}, the per-AP power constraint in~\eqref{eq:power-constraint} results in
\begin{equation} \label{eq:power-contraint:CB}
\sum\limits_{k=1}^K \eta_{mk} \gamma_{mk} \leq \frac{1}{N},~m=1,\ldots,M.
\end{equation}

\subsection{Normalized Conjugate Beamforming} \label{sec:NCB}

NCB consists in setting $\bw\NCB_{mk} = {\hat{\bg}^\ast_{mk}}/{\norm{\hat{\bg}_{mk}}}$. By inserting this into~\eqref{eq:SINR} and computing the corresponding expectations, the effective SINR per user is given by~\cite{Polegre2020} and~\eqref{eq:SINR:NCB}, shown at the bottom of this page.
In this case, the power of the beamforming gain uncertainty is equal to \addtocounter{equation}{3}
\begin{equation} \label{eq:variance:NCB}
\EX{|\mathsf{BU}\NCB_k|^2}\! =\! \Pd\! \sum\limits_{m=1}^M \eta_{mk} \left[\beta_{mk} \!+\! (N\!-\!1\!-\!\alpha^2) \gamma_{mk}\right],
\end{equation}
and the per-AP power constraint for NCB is
\begin{equation} \label{eq:power-contraint:NCB}
\sum\nolimits_{k=1}^K \eta_{mk} \leq 1,~m=1,\ldots,M.
\end{equation}

\subsection{Enhanced Normalized Conjugate Beamforming} \label{sec:ENCB}

ECB consists in setting
\begin{equation} \label{eq:ECB:precoding-vector}
\bw\ECB_{mk} = \frac{\hat{\bg}^\ast_{mk}}{\norm{\hat{\bg}_{mk}}^2}. 
\end{equation}
The reason why this normalization should enhance the normalization ${\hat{\bg}^\ast_{mk}}/{\norm{\hat{\bg}_{mk}}}$ and provide higher SE is intuitive. Consider the effective downlink channel at user $k$ and neglect the channel estimation error, 
\begin{equation} \label{eq:effective-DL-channel}
a_{kk} = \sum_{m=1}^M \sqrt{\eta_{mk}} \bg\trans_{mk} \bw_{mk} \approx \sum_{m=1}^M \sqrt{\eta_{mk}} \hat{\bg}\trans_{mk} \bw_{mk}.
\end{equation}
If $\bw_{mk} = \bw\ECB_{mk}$, then $a_{kk} \approx \sum\nolimits_{m=1}^M \sqrt{\eta_{mk}}$ \vspace*{1mm} is ideally deterministic. Conversely, if $\bw_{mk} = \bw\NCB_{mk}$ as in~\cite{Polegre2020}, then $a_{kk} \approx \sum_{m=1}^M \sqrt{\eta_{mk}} \norm{\hat{\bg}_{mk}}$, so that the fluctuations of the channel gain are reduced but not entirely equalized.  

\begin{figure*}[!b]
\normalsize
\setcounter{eqcnt2}{\value{equation}}
\setcounter{equation}{27}
\hrulefill
\begin{equation} \label{eq:SINR:ECB}
\mathsf{SINR}\ECB_k = \frac{\Pd \left(\sum\limits_{m=1}^M \sqrt{\eta_{mk} }\right)^2}{\dfrac{\Pd}{N\!-\!1} \sum\limits_{m=1}^M \eta_{mk} \left(\dfrac{\beta_{mk}}{\gamma_{mk}}\!-\!1\right)\!+\!\dfrac{\Pd}{N\!-\!1} \sum\limits_{j \neq k}^K \sum\limits_{m=1}^M \eta_{mj} \dfrac{\beta_{mk}}{\gamma_{mj}}\!+\!\Pd \sum\limits_{j \neq k}^K \Theta_{kj}~|\bvphi\herm_k \bvphi_j|^2+1},
\end{equation}
\setcounter{equation}{\value{eqcnt2}}
\end{figure*}

By inserting~\eqref{eq:ECB:precoding-vector} into~\eqref{eq:SINR} and computing the corresponding expectations, the effective SINR per user is given by~\eqref{eq:SINR:ECB} at the bottom of this page,
where \addtocounter{equation}{1}
\begin{align}
\!\Theta_{kj} \!&\triangleq\! \dfrac{N-2}{N-1} \! \sum\limits_{m=1}^M \! \eta_{mj} \frac{\beta^2_{mk}}{\beta^2_{mj}} \!+\!\!\! \sum\limits_{m=1}^M \sum\limits_{n \neq m}^M \!\! \sqrt{\eta_{mj}\eta_{nj}}\frac{\beta_{mk}\beta_{nk}}{\beta_{mj}\beta_{nj}}. \label{eq:ECB:upsilon}
\end{align}
\begin{IEEEproof}
See Appendix \textit{A}.
\end{IEEEproof}
\noindent For ECB, the power of the beamforming gain uncertainty is equal to 
\begin{equation} \label{eq:variance:ECB}
\EX{|\mathsf{BU}\ECB_k|^2} = \frac{\Pd}{N-1} \sum\limits_{m=1}^M \eta_{mk} \left(\dfrac{\beta_{mk}}{\gamma_{mk}}-1\right).
\end{equation}
By inserting $\bw_{mk} = \bw\ECB_{mk}$ into~\eqref{eq:data-transmission}, the per-AP power constraint in~\eqref{eq:power-constraint} becomes
\begin{equation} \label{eq:power-contraint:ECB}
\sum_{k=1}^K \eta_{mk}~\EX{\dfrac{1}{\norm{\hat{\bg}_{mk}}^2}} \leq 1 \implies \sum_{k=1}^K \dfrac{\eta_{mk}}{\gamma_{mk}} \leq N-1,
\end{equation}
$m = 1,\ldots,M$. Importantly, equations~\eqref{eq:SINR:ECB},~\eqref{eq:power-contraint:ECB} are defined if $N>1$. In fact, $\EX{1/{|x|^2}}$ does not converge if $x$ is a scalar circularly symmetric Gaussian random variable.

\section{Performance Analysis of Conjugate Beamforming with Downlink Training} \label{sec:performance-analysis-DT}
In this section we analyze the performance of CB when downlink training is carried out (CB-DT). 
Downlink training takes place via pilot beamforming as described in~\cite{Interdonato2019b}, and herein we extend the analysis in~\cite{Interdonato2019b} to multi-antenna APs. By conjugate beamforming the pilots, the estimation overhead is independent of the number of APs. The channel estimation overhead rather scales with $K$ and does not require any feedback from the users to the APs as the channel is reciprocal in TDD mode.

The downlink training phase is $\taudp$ samples long. This increases the pilot overhead up to $\tauup + \taudp$. Let $\sqrt{\taudp} \bpsi_k \in \C^{\taudp}$ be the downlink pilot sequence intended for user $k$, $\norm{\bpsi_k}=1$, and $\Pdp$ be the normalized SNR of the downlink pilot symbol. AP $m$ beamforms the pilots as
\begin{align} \label{eq:downlink-pilot-transmission}
\bX_{m,\mathrm{p}} = \sqrt{\taudp \Pdp} \sum\limits_{k=1}^K \sqrt{\eta_{mk}} \hat{\bg}^\ast_{mk} \bpsi\trans_k \in \C^{N \times \taudp},
\end{align}
subject to the following power constraint:
\begin{align} \label{eq:power-contraint:CBdl}
&\EX{\norm{\bX_{m,\mathrm{p}}}^2_{\mathrm{F}}} \nonumber \\ 
\quad &= \taudp \Pdp \Tr\EX{\sum\limits_{k=1}^K \sum\limits_{j=1}^K \sqrt{\eta_{mk} \eta_{mj}} \hat{\bg}^*_{mk}  \hat{\bg}\trans_{mj} \bpsi\trans_k \bpsi^\ast_j} \nonumber \\
&= \taudp \Pdp \Tr\left( \sum\limits_{k=1}^K \eta_{mk} \gamma_{mk} \bI_N \right. \nonumber \\ 
&\quad\quad\left.+\sum\limits_{k=1}^K \sum\limits_{j \neq k}^K \sqrt{\eta_{mk} \eta_{mj}} \EX{\hat{\bg}^\ast_{mk}  \hat{\bg}\trans_{mj}} \bpsi\trans_k \bpsi^\ast_j \right) \nonumber \\
&\stackrel{(a)}{=} \taudp \Pdp N \sum\limits_{k=1}^K \eta_{mk} \gamma_{mk} \leq \taudp \Pdp, 
\end{align}
where in $(a)$ we have assumed that the pilot assignment is constrained such that 
\begin{equation} \label{eq:DLpilot-assignment}
\bpsi\trans_k \bpsi^\ast_j = 0 \text{ if } \bvphi_k = \bvphi_j. 
\end{equation}
This limitation is not significant in cases of practical interest since the number of assignable uplink and downlink pilot pairs satisfying~\eqref{eq:DLpilot-assignment} is larger than the number of active users, as shown in~\cite{Interdonato2019b}. Importantly, this ensures that $\Tr \EX{\hat{\bg}^\ast_{mk} \hat{\bg}\trans_{mj}} \bpsi\trans_k \bpsi^\ast_j = 0, \forall j \neq k$. From~\eqref{eq:downlink-pilot-transmission},~ \eqref{eq:power-contraint:CBdl}, we~have
\[ \EX{\norm{\bX_{m,\mathrm{p}}}^2_{\mathrm{F}}} \leq \taudp \Pdp \implies \sum_{k=1}^K \eta_{mk} \gamma_{mk} \leq \frac{1}{N}. \]
It is observed that the data power constraint in~\eqref{eq:power-contraint:CB} and the above pilot power constraint are identical.
The pilot signal received at user $k$ is given by
\begin{align}
\by_{\mathrm{dp}, k} \!&=\! \sum\limits_{m=1}^M \bg_{mk}\trans \bX_{m,\mathrm{p}} + \bww_{\mathrm{dp}, k} \nonumber \\ 
&\!=\! \sqrt{\taudp \Pdp} \sum\limits_{j=1}^K \sum\limits_{m=1}^M \sqrt{\eta_{mj}} \bg_{mk}\trans \hat{\bg}^*_{mj}\bpsi\trans_j \!+\! \bww_{\mathrm{dp}, k}
\end{align}
$\in \C^{\taudp},$ where $\bww_{\mathrm{dp}, k}$ is an additive noise vector whose elements are \iid~$\CN(0,1)$. 

\noindent From $\by_{\mathrm{dp},k}$, user $k$ separates its channel observation as
\begin{align}
&\check{y}_{\mathrm{dp},k} = \by_{\mathrm{dp}, k} \bpsi^\ast_k \nonumber \\
&\;\!=\! \sqrt{\taudp \Pdp} a_{kk} \!+\! \sqrt{\taudp \Pdp} \sum\limits_{j\neq k}^K a_{kj} \bpsi\trans_j \bpsi^\ast_k \!+\! \bww_{\mathrm{dp}, k} \bpsi^\ast_k,
\end{align}
where $$a_{kj} \triangleq \sum\nolimits_{m=1}^M \sqrt{\eta_{mj}} \bg_{mk}\trans \hat{\bg}^\ast_{mj}$$ denotes the effective downlink channel seen by user $k$ but intended for user $j$.
Based upon $\check{y}_{\mathrm{dp},k}$, user $k$ estimates $a_{kk}$ by MMSE estimation. 
Following the same approach as in~\cite{Interdonato2019b}, we can compute the MMSE downlink channel estimate $\hat{a}_{kk}$ as well as its variance in closed form, 
\begin{align} \label{eq:kappa}
\kappa_k \!\triangleq\! \varx{\hat{a}_{kk}} \!&=\! \frac{\taudp\Pdp N^2 \left(\sum\nolimits^M_{m=1} \eta_{mk} \gamma_{mk} \beta_{mk} \right)^2}{1 \!+\! \taudp\Pdp N \sum\limits^M_{m=1} \sum\limits^K_{j=1} \eta_{mj} \gamma_{mj} \beta_{mk} |\bpsi\herm_{k} \bpsi_j|^2} \nonumber \\
\!&=\! \frac{\taudp\Pdp N^2 \varsigma_{kk}^2}{1 + \taudp\Pdp N \sum\nolimits^K_{j=1} \varsigma_{kj} |\bpsi\herm_{k} \bpsi_j|^2}. 
\end{align}
A closed-form expression for an approximate\footnote{The approximation comes from the non-Gaussian nature of $\{ a_{kj} \}$, and from~\cite{qZhang2014}. See~\cite{Interdonato2019b} for further details.} achievable downlink SE can be derived by using the so-called capacity bound with \textit{side information}~\cite[Section 2.3.5]{redbook}, and  by following the same methodology as in~\cite{Interdonato2019b}: 
\begin{equation} \label{eq:SE:dlpilots}
\mathsf{SE}_k = \xi \left(1-\frac{\tauup+\taudp}{\tauc} \right) \log_2 \left(1+\mathsf{SINR}\CBDT_k\right),
\end{equation}
with $\mathsf{SINR}\CBDT_k$ given by~\eqref{eq:SINR:CBDL:general},~\eqref{eq:SINR:CBDL} at the bottom of this page.
\begin{IEEEproof}
See Appendix \textit{B}.
\end{IEEEproof}
The SINR gain over the case where the users do not have access to the CSI is noticeable.
Compared to~\eqref{eq:SINR:CB}, in~\eqref{eq:SINR:CBDL} the coherent gain (namely the numerator of the SINR) is increased by $\Pd\kappa_k$ which is at most equal to the variance of the effective downlink channel, $\Pd N \varsigma_{kk}$. Importantly, downlink training reduces the uncertainty at the user which now can employ its CSI knowledge when performing data decoding. In fact, the channel uncertainty, $\Pd N \varsigma_{kk}$, is significantly decreased by $\Pd\kappa_k$. This residual self-interference represents the variance of the downlink channel estimation error.

\begin{figure*}[!b]
\normalsize
\setcounter{eqcnt3}{\value{equation}}
\setcounter{equation}{38}
\hrulefill
\begin{align} 
\mathsf{SINR}\CBDT_k &\approx \frac{\Pd \EX{|\hat{a}_{kk}|^2}}{\Pd \EX{|\tilde{a}_{kk}|^2} + \Pd \sum\nolimits^K_{j \neq k} \EX{|a_{kj}|^2} + 1} \label{eq:SINR:CBDL:general} \\
&= \myfrac[12pt]{\Pd N^2 \left( \sum\nolimits_{m=1}^M \sqrt{\eta_{mk}} \gamma_{mk} \right)^2+ \Pd \kappa_k}{\Pd (N\varsigma_{kk}-\kappa_k) + \Pd \sum\limits_{j \neq k}^K \left[N \varsigma_{kj} + N^2 \left|\bvphi_k\herm\bvphi_j\right|^2 \left(\sum\limits_{m=1}^M \sqrt{\eta_{mj}} \gamma_{mj} \dfrac{\beta_{mk}}{\beta_{mj}} \right)^2\, \right] + 1}. \label{eq:SINR:CBDL}
\end{align}
\setcounter{equation}{\value{eqcnt3}}
\end{figure*}

\section{Max-Min Fairness Power Control} \label{sec:MMF}
Max-min fairness (MMF) power control is an egalitarian policy that ensures maximized identical SE throughout the network. This policy perfectly suits cell-free massive MIMO which by nature guarantees a more uniform quality of service than co-located massive MIMO~\cite{Interdonato2019}. Such a sophisticated power control requires a centralized approach and solving a convex optimization problem. However, requirements on latency, computational complexity, and fronthauling load, can be relaxed by two factors: $(i)$ if relying on closed-form SE expressions, power control can operate at the large-scale fading time scale; $(ii)$ if combined with clustering, the coordination can be confined within few APs. These two aspects are key in all our implementations.

In general, the optimization problem for the MMF power control subject to per-AP power constraint can be formulated as \addtocounter{equation}{2}
\begin{subequations} \label{Problem:Max-Min:general}
\begin{align}	
  \mathop {\max}\limits_{\{\eta_{mk}\geq 0 \}} & \quad \min_{k} \mathsf{SINR}_k  \\
  \text{s.t.} &\quad \EX{\norm{\bx_m}^2} \leq \Pd, ~\forall m.
\end{align}
\end{subequations}

Next, we give rigorous formulations for the MMF optimization problem with NCB and ECB. These problems have structural similarity to that in~\cite{Ngo2017b}, hence they admit global optimal solutions that can be computed by solving a sequence of second-order cone programs.

\subsection{Problem Formulation for NCB}
Through some simple mathematical manipulations we reshape the SINR expression in~\eqref{eq:SINR:NCB}. 
Firstly, we rewrite~\eqref{eq:NCB:upsilon}~as
\begin{align} \label{eq:NCB:upsilon:max-min}
\Upsilon_{kj} &= (N-1) \sum\limits_{m=1}^M \eta_{mj} \gamma_{mj} \frac{\beta^2_{mk}}{\beta^2_{mj}} \nonumber \\
&\quad+ \alpha^2 \sum\limits_{m=1}^M \sum\limits_{n \neq m}^M \sqrt{\eta_{mj}\eta_{nj}\gamma_{mj}\gamma_{nj}}\frac{\beta_{mk}\beta_{nk}}{\beta_{mj}\beta_{nj}} \nonumber \\
&\quad + \alpha^2 \sum\limits_{m=1}^M \eta_{mj} \gamma_{mj} \frac{\beta^2_{mk}}{\beta^2_{mj}} - \alpha^2 \sum\limits_{m=1}^M \eta_{mj} \gamma_{mj} \frac{\beta^2_{mk}}{\beta^2_{mj}} \nonumber \\
&= (N-1-\alpha^2) \sum\limits_{m=1}^M \eta_{mj} \gamma_{mj} \frac{\beta^2_{mk}}{\beta^2_{mj}} \nonumber \\ 
&\quad+ \left( \alpha \sum\limits_{m=1}^M \sqrt{\eta_{mj} \gamma_{mj}} \frac{\beta_{mk}}{\beta_{mj}} \right)^2,
\end{align}
and by inserting~\eqref{eq:NCB:upsilon:max-min} into~\eqref{eq:SINR:NCB}, we obtain
\begin{equation} \label{eq:SINR:NCB:powerControl}
\mathsf{SINR}\NCB_k = \frac{\Pd \alpha^2 \left(\sum\nolimits_{m=1}^M \sqrt{\eta_{mk} \gamma_{mk}}\right)^2}{\mathsf{T1}_k + \mathsf{T2}_k + 1},
\end{equation} 
where
\begin{align}
\mathsf{T1}_k &= \Pd \sum\limits_{j=1}^K \sum\limits_{m=1}^M \eta_{mj} \left[ \beta_{mk} + (N\!-\!1\!-\!\alpha^2)\gamma_{mk}|\bvphi_k\herm \bvphi_j|^2 \right] \nonumber \\
&= \Pd \sum\limits_{j=1}^K \sum\limits_{m=1}^M \eta_{mj} \vartheta_{mkj},  \\
\vartheta_{mkj} &\triangleq \beta_{mk} + (N\!-\!1\!-\!\alpha^2)\gamma_{mk}|\bvphi_k\herm \bvphi_j|^2, \\
\mathsf{T2}_k &= \Pd \alpha^2 \sum\limits_{j \neq k}^K \left(\sum\limits_{m=1}^M \sqrt{\eta_{mj} \gamma_{mj}} \frac{\beta_{mk}}{\beta_{mj}} \right)^2|\bvphi_k\herm \bvphi_j|^2 \nonumber \\ 
&\stackrel{(a)}{=} \Pd \alpha^2 \sum\limits_{j \neq k}^K \left(\sum\limits_{m=1}^M \sqrt{\eta_{mj} \gamma_{mk}} \right)^2\!\!|\bvphi\herm_k \bvphi_j|^2.
\end{align} 
Note that $-\pi/4 < N\!-\!1\!-\!\alpha^2 < 0$, thus $\vartheta_{mkj}$ is always positive since $\beta_{mk} \geq \gamma_{mk}$. The equality $(a)$ follows from~\eqref{eq:correlated-estimates}.
By using~\eqref{eq:SINR:NCB:powerControl} the MMF optimization problem in~\eqref{Problem:Max-Min:general} can be formulated in epigraph form for NCB as
\begin{subequations} \label{Problem:Max-Min:NCB}
\begin{align}	
  \mathop {\text{maximize}}\limits_{\{\eta_{mk}\geq 0 \},~\nu} & \quad \nu  \\
  \text{s.t.} &\quad \norm{\bs_k} \leq \boldsymbol{\gamma}_{kk}\trans \bu_k,~\forall k, \label{constraint:NCB:SINR-target} \\
  			  &\quad \norm{\bu'_m} \leq \sqrt{\Pd},~\forall m, \label{constraint:NCB:power}  			  
\end{align}
\end{subequations}
where $\nu$ is a new variable which represents the minimum SINR among the users that has to be maximized, and:
\begin{itemize}
\item $\bU\!=\![\bu_1,\ldots,\bu_K]$, $\bu_k\!=\!\sqrt{\Pd}\left[\sqrt{\eta_{1k}}, \ldots, \sqrt{\eta_{Mk}}  \right]\trans$, and $\bu'_m$ is the $m$th row of $\bU$;
\item $\boldsymbol{\gamma}_{kj} = \alpha~|\bvphi_k\herm \bvphi_j|~\left[\sqrt{\gamma_{1k}}, \ldots, \sqrt{\gamma_{Mk}}  \right]\trans$;
\item $\bs_k = \sqrt{\nu}\cdot\left[\bv_{k}\trans \bI_{-k}, \norm{\bb_{k1} \circ \bu_1}, \ldots, \norm{\bb_{kK} \circ \bu_K}, 1  \right]\trans$;
\item $\bv_{k} \triangleq \left[ \boldsymbol{\gamma}\trans_{k1} \bu_1, \ldots, \boldsymbol{\gamma}\trans_{kK} \bu_K \right]\trans$;
\item $\bI_{-k}$ is a $K\times(K-1)$ matrix obtained from $\bI_k$ with the $k$th column removed;
\item $\bb_{kj} = \left[\sqrt{\vartheta_{1kj}}, \ldots, \sqrt{\vartheta_{Mkj}}  \right]\trans$.
\end{itemize}
The constraints~\eqref{constraint:NCB:SINR-target},~\eqref{constraint:NCB:power} are jointly second-order cones with respect to the power control coefficients and $\nu$. If $\nu$ is fixed, then~\eqref{Problem:Max-Min:NCB} is convex, and the globally optimal solution can be obtained by using interior-point methods. 
The globally optimal solution to~\eqref{Problem:Max-Min:NCB}  can be obtained in polynomial time via the \textit{bisection method}~\cite{Boyd2004} by solving a sequence of convex (more specifically, second-order cone) feasibility problems.
A detailed description of the bisection search algorithm for the MMF power control and a more general formulation for the MMF optimization problem can be found, for example, in~\cite{Ngo2017b},~\cite{Chien2016} and~\cite[Section 7.1.1]{massivemimobook}, respectively.

\begin{figure*}[!b]
\normalsize
\setcounter{eqcnt4}{\value{equation}}
\setcounter{equation}{48}
\hrulefill
\begin{align} \label{eq:SINR:ECB:powercontrol}
\mathsf{SINR}\ECB_k &= \frac{\Pd \left(\sum\nolimits_{m=1}^M \sqrt{\eta_{mk} }\right)^2}{\dfrac{\Pd}{N\!-\!1} \sum\limits_{j=1}^K \sum\limits_{m=1}^M \eta_{mj} \left( \dfrac{\beta_{mk}}{\gamma_{mj}}\!-\!\dfrac{\beta^2_{mk}}{\beta^2_{mj}}|\bvphi\herm_k \bvphi_j|^2\right)\!+\!\Pd \sum\limits_{j \neq k}^K \! \left(\sum\limits_{m=1}^M \sqrt{\eta_{mj}}\dfrac{\beta_{mk}}{\beta_{mj}}\right)^2\!\!|\bvphi\herm_k \bvphi_j|^2\!+\!1} \nonumber \\
&= \frac{\Pd \left(\sum\nolimits_{m=1}^M \sqrt{\eta_{mk} }\right)^2}{\Pd \sum\limits_{j=1}^K \sum\limits_{m=1}^M \eta_{mj} \varrho_{mkj} \!+\!\Pd \sum\limits_{j \neq k}^K \! \left(\sum\limits_{m=1}^M \sqrt{\eta_{mj}}\dfrac{\beta_{mk}}{\beta_{mj}}\right)^2\!\!|\bvphi\herm_k \bvphi_j|^2\!+\!1},
\end{align}
\setcounter{equation}{\value{eqcnt4}}
\end{figure*}

\subsection{Problem Formulation for ECB}
Similarly, for ECB we first realize that~\eqref{eq:ECB:upsilon} can be written~as
\begin{align}
\Theta_{kj} &= \dfrac{N-2}{N-1}\sum\limits_{m=1}^M \eta_{mj} \frac{\beta^2_{mk}}{\beta^2_{mj}} + \sum\limits_{m=1}^M \sum\limits_{n \neq m}^M \sqrt{\eta_{mj}\eta_{nj}}\frac{\beta_{mk}\beta_{nk}}{\beta_{mj}\beta_{nj}} \nonumber \\
&\quad + \sum\limits_{m=1}^M \eta_{mj} \frac{\beta^2_{mk}}{\beta^2_{mj}} - \sum\limits_{m=1}^M \eta_{mj} \frac{\beta^2_{mk}}{\beta^2_{mj}} \nonumber \\
&= \left(\dfrac{N-2}{N-1}-1\right) \sum\limits_{m=1}^M \eta_{mj} \frac{\beta^2_{mk}}{\beta^2_{mj}} + \left(\sum\limits_{m=1}^M \sqrt{\eta_{mj}}\frac{\beta_{mk}}{\beta_{mj}}\right)^2 \nonumber \\
&= \left(\sum\limits_{m=1}^M \sqrt{\eta_{mj}}\frac{\beta_{mk}}{\beta_{mj}}\right)^2 - \dfrac{1}{N-1} \sum\limits_{m=1}^M \eta_{mj} \frac{\beta^2_{mk}}{\beta^2_{mj}}.
\end{align}
As a result, the SINR expression in~\eqref{eq:SINR:ECB} is given by~\eqref{eq:SINR:ECB:powercontrol} at the bottom of this page,
where we have defined $$\varrho_{mkj} \triangleq \dfrac{1}{{N\!-\!1}}\left( \dfrac{\beta_{mk}}{\gamma_{mj}}\!-\!\dfrac{\beta^2_{mk}}{\beta^2_{mj}}|\bvphi\herm_k \bvphi_j|^2\right),$$ which is always positive since $\beta_{mk} \geq \gamma_{mk}$, and $N>1$.
The MMF optimization problem in~\eqref{Problem:Max-Min:general} can be formulated in epigraph form for ECB as \addtocounter{equation}{1}
\begin{subequations} \label{Problem:Max-Min:ECB}
\begin{align}	
  \mathop {\text{maximize}}\limits_{\{\eta_{mk}\geq 0 \},~\nu} & \quad \nu  \\
  \text{s.t.} &\quad \norm{\bs_k} \leq \boldsymbol{1}\trans \bu_k,~\forall k, \\
  			  &\quad \norm{\hat{\boldsymbol{\gamma}}'_m \circ \bu'_m} \leq \sqrt{\Pd(N-1)},~\forall m,
\end{align}
\end{subequations}
where $\nu$, $\bu_k$ and $\bu'_m$ are defined as in~\eqref{Problem:Max-Min:NCB}, $\boldsymbol{1}$ denotes the $M$-dimensional vector of ones, and: 
\begin{itemize}
\item $\bs_k = \sqrt{\nu}\cdot\left[\bv_{k}\trans \bI_{-k}, \norm{\bz_{k1} \circ \bu_1}, \ldots, \norm{\bz_{kK} \circ \bu_K}, 1  \right]\trans$;
\item $\bv_{k} \triangleq \left[ \bb\trans_{k1} \bu_1, \ldots, \bb\trans_{kK} \bu_K \right]\trans$;
\item $\hat{\boldsymbol{\gamma}}'_m$ is the $m$th row of $\hat{\boldsymbol{\Gamma}}$, $\hat{\boldsymbol{\Gamma}}\!=\![\hat{\boldsymbol{\gamma}}_1,\ldots,\hat{\boldsymbol{\gamma}}_K]$ and \\ $\hat{\boldsymbol{\gamma}}_k = \left[\gamma_{1k}^{-1/2}, \ldots, \gamma_{Mk}^{-1/2}  \right]\trans$; 
\item $\bb_{kj} = |\bvphi_k\herm \bvphi_j|\left[\dfrac{\beta_{1k}}{\beta_{1j}}, \ldots, \dfrac{\beta_{Mk}}{\beta_{Mj}}\right]\trans$; \vspace*{2mm}
\item $\bz_{kj} = \left[\sqrt{\varrho_{1kj}},\ldots,\sqrt{\varrho_{Mkj}} \right]\trans$.
\end{itemize}

\subsection{Problem Formulation for CB with Downlink Training}
The formulation of the power optimization problem for the CB scheme with downlink training and multi-antenna APs follows that in~\cite[Section IV-A]{Interdonato2019b}, with a simple adjustment: $\gamma_{mk}$ must be replaced with $N\gamma_{mk}$ everywhere in the problem formulation.
Compared to NCB and ECB, the MMF power control for CB with downlink training does not rigorously provide identical SINRs as it results from a sequential convex approximation and thereby the problem solutions are sub-optimal~\cite{Interdonato2019b}.

\section{Simulation Results} \label{sec:results}

In this section, we compare the performance of the precoding schemes discussed in~\Secref{sec:performance-analysis} and~\ref{sec:performance-analysis-DT} by presenting the results of our simulations. Next, we introduce the simulation scenario and the adopted settings.

\subsection{Simulation Scenario and Settings}
In our simulations we consider a nominal area of $D \times D$ squared meters, wherein APs and users are uniformly located at random. A wraparound technique is used to simulate a cell-free network. 
A random realization of AP and user locations determines a set of large-scale fading coefficients and constitutes a snapshot of the network. For a network snapshot the achievable downlink SEs are computed, according to the expressions presented in~\Secref{sec:performance-analysis} and~\ref{sec:performance-analysis-DT}. The cumulative distribution function (CDF) of the SE is obtained over many network snapshots.

We adopt the 3GPP Urban Microcell pathloss model defined in \cite[Table B.1.2.1-1]{LTE2017} as
\begin{equation}
\label{eq:path-loss-model}
\mathsf{PL}_{mk}~\text{[dB]} = -30.5-36.7 \log_{10} \left(\frac{d_{mk}}{1~\text{m}} \right),
\end{equation}
where $d_{mk}$ is the distance (in three dimensions) between AP $m$ and user $k$, and assuming a 2~GHz carrier frequency. 
We also consider log-normal shadow fading with standard deviation $\sigma_\text{sh}$ and spatial correlations both at the APs and the users. More specifically, let $q_{mk}\sim\mathcal{N}(0,1)$ be defined as
\begin{equation}
\label{eq:shadowing-model}
q_{mk} = \sqrt{\epsilon}~a_m + \sqrt{1-\epsilon}~b_k,
\end{equation}  
where $a_m \sim \mathcal{N}(0,1)$ and $b_k \sim \mathcal{N}(0,1)$ are independent random variables capturing the shadow fading effects from AP $m$ to all the users and from user $k$ to all the APs, respectively. The parameter $0<\epsilon<1$ weighs these effects.
The shadow fading spatial correlations are thus modeled as~\cite{LTE2017} 
\begin{equation}
\EX{a_m a_n} = 2^{-(d^{\mathsf{~\!\!AP}}_{mn} \left/\right. 9\text{ m})}, \qquad \EX{b_k b_j} = 2^{-(d^{\mathsf{~\!\!UE}}_{kj} \left/\right. 9\text{ m})},
\end{equation}    
where $d^{\mathsf{~\!\!AP}}_{mn}$ is the distance between AP $m$ and  AP $n$, $d^{\mathsf{~\!\!UE}}_{kj}$ is the distance between user $k$ and user $j$, and 9 meters is the decorrelation distance. Pathloss and log-normal shadow fading enter into the large-scale fading coefficients as
\begin{equation}
\label{eq:beta}
\beta_{mk} = \mathsf{PL}_{mk} \cdot 10^{~\sigma_\text{sh}q_{mk} / 10}.
\end{equation}
Unless otherwise stated, we use the following simulation settings: $D = 500$ m, $\sigma_\text{sh}=4$~dB, $\epsilon = 0.5$, AP height 10 m, user height 1.5 m, channel bandwidth $B=20$ MHz, antenna gains 0 dBi. The TDD coherence block is partitioned equally between uplink and downlink, so $\xi=0.5$, and it is $\tc = 200$ samples long, resulting from a coherence bandwidth of 200 kHz and a coherence time of 1 ms.
The maximum transmit power per AP and per user is $200$ mW and $100$~mW, respectively. This is normalized by the noise power, $n^{\mathsf{(dBm)}}_\mathsf{p}=-92$ dBm, to obtain
\begin{align}
\Pd \text{ [dBm]} &= \Pdp \text{ [dBm]} = 10\log_{10}(200) - n^{\mathsf{(dBm)}}_\mathsf{p}, \\
\Pu \text{ [dBm]} &= 10\log_{10}(100) - n^{\mathsf{(dBm)}}_\mathsf{p}.
\end{align}  

In all our simulations, we also consider the largest-large-scale-fading-based AP selection~\cite{Ngo2018a}, according to which an AP participates in the service of user $k$ if its channel is sufficiently strong, and more specifically if $\beta_{mk}$ belongs to the user-$k$-specific AP cluster $\mathcal{A}_k$ which satisfies   
\begin{equation}
\sum\limits_{m=1}^{|\mathcal{A}_k|} \frac{\bar{\beta}_{mk}}{\sum\nolimits_{n=1}^M \beta_{nk}} \geq 95\%,
\end{equation}
where $|\mathcal{A}_k|$ is the cardinality of the set $\mathcal{A}_k$ with $\min(|\mathcal{A}_k|) = 10$, and  $\{\bar{\beta}_{1k}, \ldots, \bar{\beta}_{M,k}\}$ are the large-scale fading coefficients sorted in descending order.

To emphasize that the power control coefficients of the considered precoding schemes are subject to different per-AP power constraints, we denote by $\{\eta\CB_{mk}\}$, $\{\eta\NCB_{mk}\}$ and $\{\eta\ECB_{mk}\}$ the power control coefficients related to CB, NCB and ECB, respectively. Note that the power constraints for CB-DT and CB are identical, hence we use $\{\eta\CB_{mk}\}$ to denote the coefficients for CB-DT too. \Tableref{tab:power-constraints} reports the per-AP power constraints for the considered precoding schemes.
\begin{table}[!t] \centering \normalsize
\caption{Per-AP power constraints for different precoding schemes.}
\begin{tabular}{@{}c|c|c@{}}
\toprule
CB, CB-DT &  NCB & ECB  \\ \midrule
$N \sum\limits_{k=1}^K \eta\CB_{mk} \gamma_{mk} \leq 1$
 & $\sum\limits_{k=1}^K \eta\NCB_{mk} \leq 1$ & $\dfrac{1}{N-1} \sum\limits_{k=1}^K \dfrac{\eta\ECB_{mk}}{\gamma_{mk}} \leq 1$ \\
\bottomrule
\end{tabular}
\label{tab:power-constraints}
\end{table}  
As an alternative to the MMF power control described in~\Secref{sec:MMF}, we consider a heuristic distributed power control, also knows as \textit{maximal-ratio}~\cite{Ngo2017b}, which consists in setting the power control coefficients as
\begin{align}
\eta\CB_{mk} &= \frac{1}{N \sum\nolimits_{j=1}^K \gamma_{mj}}, \label{eq:eta:CB} \\
\eta\NCB_{mk} &= N \gamma_{mk} \eta\CB_{mk}, \label{eq:eta:NCB} \\
\eta\ECB_{mk} &= (N-1) \gamma_{mk} \eta\NCB_{mk}. \label{eq:eta:ECB}
\end{align} 
With these settings the respective power constraints hold with equality, and thereby each AP spends all the available transmit power. Maximal-ratio power control is an ``opportunistic'' policy whereby more power is allocated to the users with better channel conditions.

Finally, since $\tauup < K$, we assume that the uplink pilot sequences are assigned at random and reused throughout the network. When downlink training is performed, the downlink pilot sequences are assigned as proposed in~\cite{Interdonato2019b} to satisfy the condition in~\eqref{eq:DLpilot-assignment}.

\subsection{Performance Evaluation} \label{sec:evaluation}

The first aspect we focus on is the impact of the precoding normalization on the performance.
In this regard, we look at two metrics: $(i)$ the power of the self-interference as share of the desired signal power which is also known as \textit{coherent} gain; $(ii)$ the power of the inter-user interference as share of the coherent gain.

If downlink training is not performed, the self-interference corresponds to the beamforming gain uncertainty introduced in~\Secref{sec:performance-analysis}. Hence, the first metric is simply
\begin{equation} \label{eq:metric1}
\frac{\EX{|\mathsf{BU}_k|^2}}{|\mathsf{DS}_k|^2},
\end{equation}
where $\EX{|\mathsf{BU}_k|^2}$ is equal to $\Pd N \varsigma_{kk}$ for CB, and given by~\eqref{eq:variance:NCB} and~\eqref{eq:variance:ECB} for NCB and ECB, respectively. The coherent gain is given by the term in the numerator of the SINR expressions in~\eqref{eq:SINR:CB},~\eqref{eq:SINR:NCB} and~\eqref{eq:SINR:ECB}, respectively.

Importantly, the metric~\eqref{eq:metric1} gives a meaningful measure of the channel hardening, since it is a normalized variance of the effective downlink channel for user $k$. The smaller this variance is, the more the channel hardens. 
Conversely, for the CB scheme with downlink training (CB-DT) the self-interference corresponds to the variance of the downlink channel estimation error, $\Pd \EX{|\tilde{a}_{kk}|^2} = \Pd N \varsigma_{kk} - \Pd \kappa_k$, i.e., the first term of the denominator in~\eqref{eq:SINR:CBDL}, the coherent gain is the term in the numerator of~\eqref{eq:SINR:CBDL}, namely $\Pd \EX{|\hat{a}_{kk}|^2}$, and the inter-user interference is the second term of the denominator in~\eqref{eq:SINR:CBDL}, corresponding to $\Pd \sum_{j \neq k}^K \EX{|a_{kj}|^2}$.  

\Tableref{tab:formulas-comparison} summarizes the closed-form expressions of coherent gain, self-interference and inter-user interference for all the considered precoding schemes.
\begin{table*}[!t] \centering \normalsize
\caption{Explicit closed-form expressions of the expectations in~\eqref{eq:SINR} and~\eqref{eq:SINR:CBDL:general}.}
\begin{tabular}{@{}l|c|c@{}}
\toprule
 &  Coherent Gain &  Self-Interference  \\ \midrule
CB & $\Pd N^2 \left( \sum\limits_{m=1}^M \sqrt{\eta\CB_{mk}} \gamma_{mk} \right)^2$  & $\Pd N\sum\limits_{m=1}^M \eta\CB_{mk} \beta_{mk} \gamma_{mk}$  \\
NCB & $\Pd \alpha^2 \left(\sum\limits_{m=1}^M \sqrt{\eta\NCB_{mk} \gamma_{mk}}\right)^2$  & $\Pd \sum\limits_{m=1}^M \eta\NCB_{mk} \beta_{mk}\!+\!\Pd (N\!-\!1\!-\!\alpha^2) \sum\limits_{m=1}^M \eta\NCB_{mk}\gamma_{mk}$   \\
ECB & $\Pd \left(\sum\limits_{m=1}^M \sqrt{\eta\ECB_{mk} }\right)^2$  & $\dfrac{\Pd}{N\!-\!1} \sum\limits_{m=1}^M \eta\ECB_{mk} \left( \dfrac{\beta_{mk}}{\gamma_{mk}}\!-\!1\right)$  \\
CB-DT & $\Pd N^2 \left( \sum\limits_{m=1}^M \sqrt{\eta\CB_{mk}} \gamma_{mk} \right)^2\!\!\!+ \Pd \kappa_k$  & $\Pd N\sum\limits_{m=1}^M \eta\CB_{mk} \beta_{mk} \gamma_{mk} - \Pd\kappa_k$  \\ \midrule
& \multicolumn{2}{c}{Inter-user Interference} \\ \midrule
CB, CB-DT & \multicolumn{2}{c}{$\Pd N \sum\limits_{j \neq k}^K \left[ \sum\limits_{m=1}^M \eta\CB_{mj} \beta_{mk} \gamma_{mj}\!+\!N \left(\sum\limits_{m=1}^M \sqrt{\eta\CB_{mj}} \gamma_{mj} \dfrac{\beta_{mk}}{\beta_{mj}} \right)^2\!\!\left|\bvphi_k\herm\bvphi_j\right|^2 \right]$} \\
NCB & \multicolumn{2}{c}{$\Pd \sum\limits_{j \neq k}^K \left[ \sum\limits_{m=1}^M \! \eta\NCB_{mj} \left(\beta_{mk}\!+\!(N\!-\!1\!-\!\alpha^2)\gamma_{mk}|\bvphi\herm_k \bvphi_j|^2\right)\!+\!\bigg(\!\alpha\!\!\sum\limits_{m=1}^M \!\! \sqrt{\eta\NCB_{mj} \gamma_{mk}} \bigg)^2\!\!|\bvphi\herm_k \bvphi_j|^2 \right]$} \\
ECB & \multicolumn{2}{c}{$\Pd \sum\limits_{j \neq k}^K \left[ \dfrac{1}{{N\!-\!1}} \sum\limits_{m=1}^M \! \eta\ECB_{mj} \! \left( \dfrac{\beta_{mk}}{\gamma_{mj}}\!-\!\dfrac{\beta^2_{mk}}{\beta^2_{mj}}|\bvphi\herm_k \bvphi_j|^2\right) \!+\! \left(\sum\limits_{m=1}^M \sqrt{\eta\ECB_{mj}}\dfrac{\beta_{mk}}{\beta_{mj}}\right)^2\!\!|\bvphi\herm_k \bvphi_j|^2 \right]$} \\
\bottomrule
\end{tabular}
\label{tab:formulas-comparison}
\end{table*}

\begin{figure}[!t]
\centering
\includegraphics[width=.85\linewidth]{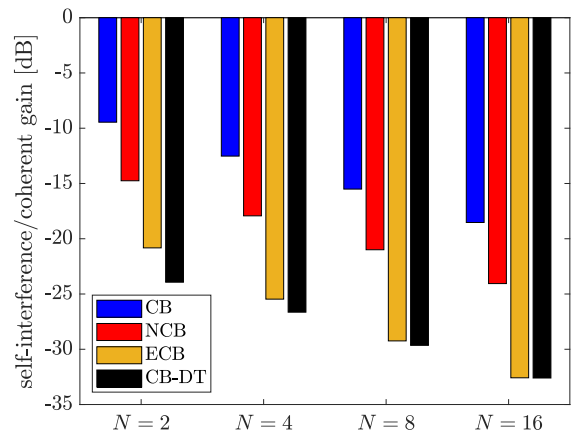}
\caption{Average self-interference to coherent gain ratio in dB, for different numbers of antennas per AP. In this simulation: $M = 200$, $K = 40$, $\tauup = \taudp = 20$ and maximal-ratio power control.}
\label{fig:fig1}
\end{figure}

In~\Figref{fig:fig1}, we show the average (over many network snapshots) ratio of the self-interference to the coherent gain for different numbers of antennas per AP, $N = \{2,4,8,16\}$.
In this simulation, we adopt the following settings: $M = 200$ APs, $K = 40$ users, $\tauup = \taudp = 20$ pilots and maximal-ratio power control. The results in~\Figref{fig:fig1} demonstrate the outstanding ability of ECB to boost the channel hardening compared to NCB and CB, reducing the normalized beamforming gain uncertainty by at least 5 dB and 10 dB, respectively. In addition, ECB is able to provide almost the same amount of self-interference to coherent gain ratio as CB-DT (the gap reduces as the number of antennas per AP increases), which tells us that the CSI acquisition at the user, although reducing the uncertainty about the channel, does not provide an added value in terms of performance. 

\begin{figure}[!t]
\centering
\includegraphics[width=.85\linewidth]{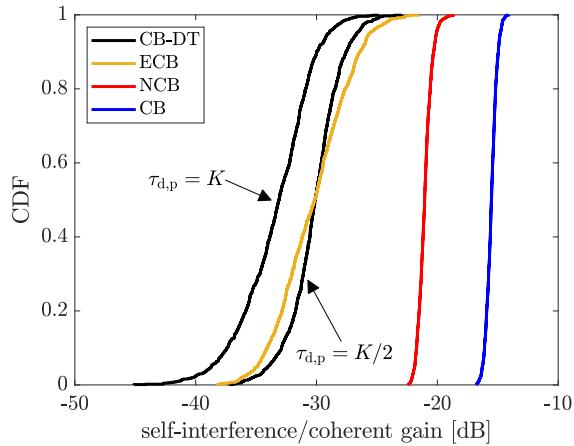}
\caption{CDF of the self-interference to coherent gain ratio in dB, for  $N=8$. The settings are identical to those in~\Figref{fig:fig1}.}
\label{fig:fig2}
\end{figure}

Certainly, the variance of the downlink channel estimation error depends on the level of downlink pilot contamination which in turn depends on the number of orthogonal downlink pilots. The self-interference for CB-DT is minimized if each user is assigned a unique orthogonal downlink pilot (no pilot reuse). Hence, it is interesting to look at how far the performance of ECB is from that ideal case. In~\Figref{fig:fig2} we show the CDF of the self-interference to coherent gain ratio for $N = 8$ whose mean value has been shown in~\Figref{fig:fig1}.     
In these results, we include the CB-DT with $\taudp = K$, i.e., no downlink pilot reuse. Compared to this ideal case, ECB only loses uniformly around 3 dB. However, this additional pilot overhead significantly reduces the SE of CB-DT, as shown in~\eqref{eq:SE:dlpilots}.

Another important aspect to look at is how the considered precoding schemes tackle the inter-user interference and how much coherent gain they achieve. None of the considered schemes provides interference suppression by nature, and the amount of interference in the network remains essentially the same regardless of the variant of CB that is adopted. In fact, the transmit power of any AP is, in any case, equal to $\Pd$ which is ensured by performing the maximal-ratio power control scheme described by~\eqrefs{eq:eta:CB}{eq:eta:ECB}. 

\begin{figure}[!t]
\centering
\includegraphics[width=.85\linewidth]{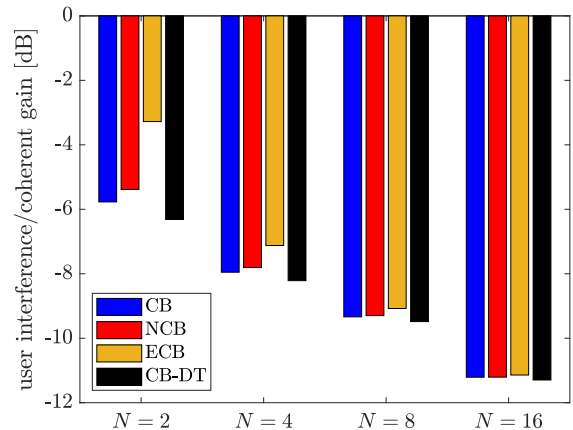}
\caption{Average inter-user interference to coherent gain ratio in dB, for different numbers of antennas per AP. The simulation settings are identical to those in~\Figref{fig:fig1}.}
\label{fig:fig3}
\end{figure}

From the same set of simulations used so far, we now show in~\Figref{fig:fig3} the power of the inter-user interference as share of the coherent gain, for different setups: $N = \{2,4,8,16\}$. From~\Figref{fig:fig3} we observe that the inter-user interference to coherent gain ratio decreases with the number of antennas per AP thanks to an increasing coherent gain.
Interestingly, ECB performs poorly when $N=2$. By substituting~\eqrefs{eq:eta:CB}{eq:eta:ECB} into the expressions of the coherent gain in Table III, we indeed observe that the coherent gain is proportional to $N$ and $N-1$ for CB and ECB, respectively. If $N = 2$, then the coherent gain of ECB is half the coherent gain of CB. This 3 dB loss with respect to CB can be observed in~\Figref{fig:fig3}. Clearly, this gap vanishes as $N$ grows.

Importantly,~\Figref{fig:fig3} tells us that the precoding normalization has a significant impact only on the self-interference and in absence of CSI at the users. Hence, we can argue that any precoding normalization along with downlink training would yield negligible benefits compared to CB-DT. We will use this important consideration to draw general conclusions about the usefulness of the downlink training.

\begin{figure}
     \centering \vspace*{-4mm}
     \subfloat[][CDF of the achievable gross SE per user]{\includegraphics[width=.95\linewidth]{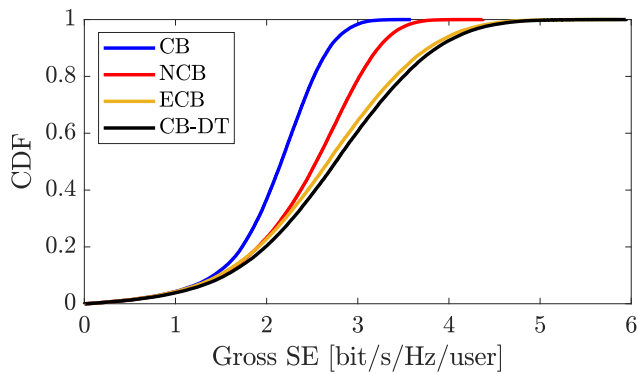}\label{fig:fig4a}} \\ \vspace*{-2mm}
     \subfloat[][CDF of the achievable (net) SE per user]{\includegraphics[width=.95\linewidth]{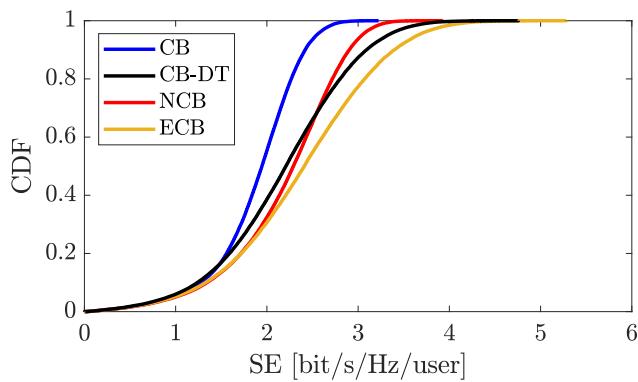}\label{fig:fig4b}}
     \caption{SE with maximal-ratio power control. The simulation settings are identical to those in~\Figref{fig:fig1}. Here, we consider $N=8$.}
     \label{fig:fig4}
\end{figure}

In~\Figref{fig:fig4}, we present the CDF of the achievable SE for the considered precoding schemes. The simulation settings are the same used so far, and we consider $N = 8$ and maximal-ratio power control. \Figref{fig:fig4a} shows the \textit{gross} SE, namely~\eqref{eq:SE} and~\eqref{eq:SE:dlpilots} without the pre-log factor capturing the pilot estimation overhead. By doing so, we want to emphasize how ECB uniformly performs tightly close to CB-DT\footnote{Strictly speaking, CB-DT is not an ``upper-bound'', but constitutes a realistic performance benchmark considering distributed conjugate beamforming schemes relying only on local channel estimates. For instance, the modified CB proposed in~\cite{Attarifar2019} can achieve signal-to-interference ratio (SIR) values close to a genie-aided receiver, but requires CSI exchange among the APs.}, regardless of the pilot overhead. \Figref{fig:fig4b} shows how the additional pilot overhead negatively affects the SE, making ECB the most desirable precoding scheme.

\begin{figure}
     \centering \vspace*{-2mm}
     \subfloat[][CDF of the achievable min SE per user with $\tauc = 200$.]{\includegraphics[width=.95\linewidth]{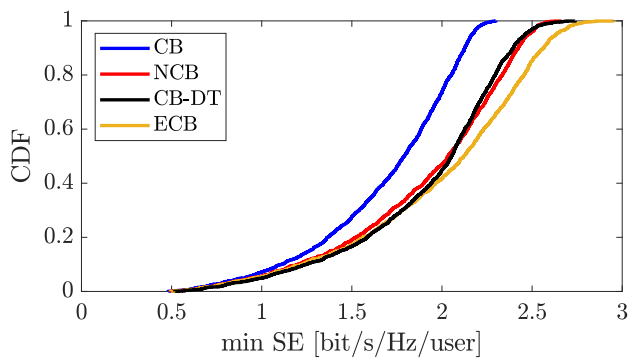}\label{fig:fig5a}} \\ \vspace*{-2mm}
     \subfloat[][CDF of the achievable min SE per user with $\tauc = 100$.]{\includegraphics[width=.95\linewidth]{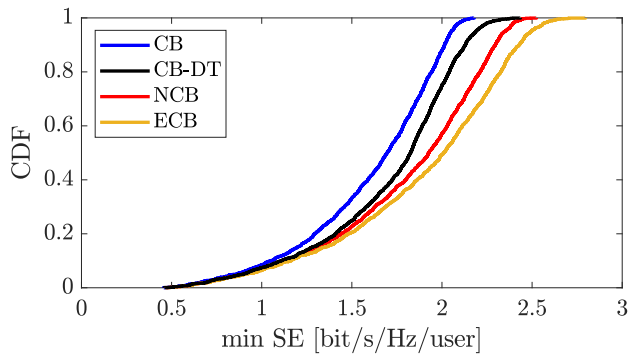}\label{fig:fig5b}} 
     \caption{SE with MMF power control. Here $M = 100$ APs, $K = 20$ users, $\tauup = \taudp = 10$ pilots and $D = 250 $ m.}
     \label{fig:fig5}
\end{figure}

ECB outperforms NCB, especially at high percentiles but this is due to the opportunistic nature of the maximal-ratio power control which prioritizes the users with stronger channel. In fact, if we perform MMF power control as described in~\Secref{sec:MMF} a small gain can be observed for the minimum SE per user (see~\Figref{fig:fig5a}).
The reason why this gain is relatively small is intuitive: the power control coefficients $\{\eta_{mk}\}$ are part of the effective downlink channel gain, thus when optimized, they act on the normalization in different ways (whereas in maximal-ratio power control they are proportional to $\gamma_{mk}$) in order to maximize the minimum SE. Hence, MMF power control tends to reduce the differences between NCB and ECB. 
An additional reason why there is very little difference in SE at low percentiles between the different methods is that the inter-user interference to coherent gain ratio, which is basically the same for all the schemes for $N = 8$, is dominant over the self-interference to coherent gain ratio, as shown in~\Figref{fig:fig1} and~\Figref{fig:fig3}, hence the gain provided by ECB (and CB-DT) in terms of self-interference becomes negligible.
CB-DT benefits from optimal power control for an additional aspect: downlink pilots are beamformed and power-controlled by the same power control coefficients $\{\eta\CB_{mk}\}$ used for the data transmission. Hence, MMF power control adjusts the power levels of the downlink pilot transmissions to reduce the pilot contamination and achieve the target SINR. However, as we can see from~\Figref{fig:fig5a}, this is not sufficient to outperform ECB, which is the preferable scheme.

The settings adopted for the simulations in~\Figref{fig:fig5a} consist of: $M = 100$ APs equipped with $N = 8$ antennas, $K = 20$ users, $\tauup = \taudp = 10$ pilots, $\tauc=200$ and $D = 250 $ m. 
Being quite sensitive to the pilot overhead, CB-DT is significantly affected by the length of the coherence block. 
If we shrink the coherence block to $\tauc = 100$, then the resulting min SE of CB-DT, shown in~\Figref{fig:fig5b}, degrades more rapidly compared to the other downlink-training-free schemes.

\begin{figure}[!t]
\centering
\includegraphics[width=.8\linewidth]{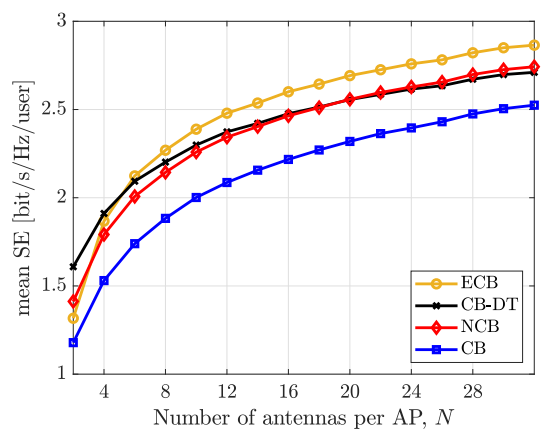}
\caption{Mean SE as the number of antennas per AP varies. The simulation settings are identical to those in~\Figref{fig:fig4}.}
\label{fig:fig6}
\end{figure}

\begin{figure}[!t]
\centering
\includegraphics[width=.8\linewidth]{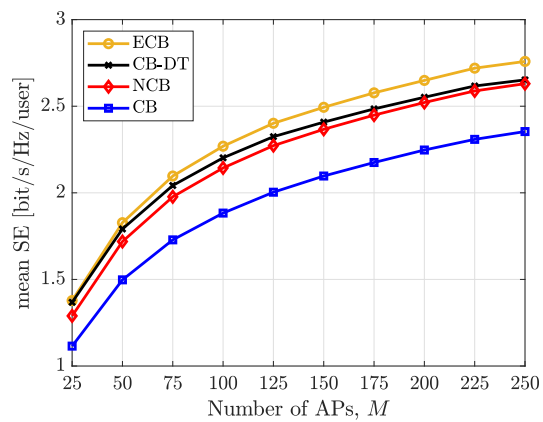}
\caption{Mean SE as the number of APs varies. The simulation settings are identical to those in~\Figref{fig:fig4}.}
\label{fig:fig7}
\end{figure}

Finally, we investigate how the mean SE varies with the number of antennas per APs (\Figref{fig:fig6}) respectively the number of APs (\Figref{fig:fig7}).
Increasing the number of antennas per AP always boosts the channel hardening~\cite{ZChen2018}, and this makes downlink training unnecessary. In fact,~\Figref{fig:fig6} shows that CB-DT is preferable for very small values of $N$, while ECB becomes the best scheme with $N\geq 6$, and the gap between ECB and CB-DT increases with $N$. Importantly, the ability of NCB to help the effective downlink channel to harden is inferior compared to ECB, and NCB performs at most equally as CB-DT.
Increasing the number of APs, $M$, is not as crucial as increasing $N$ for the channel to harden~\cite{ZChen2018}. \Figref{fig:fig7} shows that ECB uniformly outperforms all the other schemes and the performance improvement increases with $M$, although slower than it would increase with $N$. Interestingly, NCB does not outperform CB-DT despite $N=8$ and the additional overhead that penalizes CB-DT.

In summary, ECB offers a better support for the channel to harden than NCB and its excellent performance makes downlink training unnecessary in the considered scenarios. 

\section{Conclusion}
In this work, we have studied a variant of the conjugate beamforming scheme, dubbed enhanced normalized conjugate beamforming (ECB), for cell-free massive MIMO systems with multi-antenna APs. The ECB precoding vector consists of the conjugate of the channel estimate normalized by its squared norm. This normalization term is the core of our contribution and leads to many benefits. Firstly, this precoding normalization helps, more than any other normalization proposed in the literature, to achieve channel hardening, i.e., to make the effective downlink channel gain nearly deterministic. This, in turn, makes data decoding methods based on the channel statistics more reliable. 
Secondly, we have demonstrated that ECB with statistical CSI knowledge at the users can provide better downlink spectral efficiency than conjugate beamforming with pilot-based downlink training, even with relatively small numbers of APs and antennas per AP. Since the precoding normalization significantly affects only the self-interference due to the user's channel uncertainty, then it would not appreciably increase the performance in the presence of downlink training. We can thereby conclude that ECB might render the downlink pilots unnecessary from a performance viewpoint. This conclusion does not contrast with our previous work~\cite{Interdonato2019b} but rather closes the loop. In fact, the performance gap between the conventional conjugate beamforming with and without downlink training is considerable, even with multi-antenna APs, but can be filled up by the precoding normalization herein proposed.

\vspace*{-2mm}
\appendix
\subsection{Proof of~\eqref{eq:SINR:ECB}}
Next, we include the computation in closed form of the expectations needed to derive the SINR expression in~\eqref{eq:SINR:ECB}.
Let $\bg_{mk} = \hat{\bg}_{mk} + \tilde{\bg}_{mk} \in \C^N$ and $\hat{\bg}_{mk}$ be independent of $\tilde{\bg}_{mk}$. Moreover, $\bg_{mk}\!\sim\!\CN(\bzero,\beta_{mk}\bI_N)$, $\hat{\bg}_{mk}\!\sim\!\CN(\bzero,\gamma_{mk}\bI_N)$ and $\tilde{\bg}_{mk}\!\sim\!\CN(\bzero,(\beta_{mk}-\gamma_{mk})\bI_N)$. It holds that, 
\begin{align}
&\EX{\frac{ \bg_{mk}\trans \hat{\bg}_{mk}^\ast}{\norm{\hat{\bg}_{mk}}^2}}\! = \! \EX{\EX{\frac{\hat{\bg}_{mk}\trans \hat{\bg}_{mk}^\ast}{\norm{\hat{\bg}_{mk}}^2}+\frac{\tilde{\bg}_{mk}\trans \hat{\bg}_{mk}^\ast}{\norm{\hat{\bg}_{mk}}^2}\bigg\rvert \hat{\bg}_{mk}}} \!=\! 1, \\ \nonumber
&\EX{\frac{ |\bg_{mk}\trans \hat{\bg}_{mk}^\ast|^2}{\norm{\hat{\bg}_{mk}}^4}} = \EX{\frac{ |(\hat{\bg}_{mk} + \tilde{\bg}_{mk})\trans \hat{\bg}_{mk}^\ast|^2}{\norm{\hat{\bg}_{mk}}^4}} \nonumber \\
&\quad= 1+\EX{|\tilde{\bg}_{mk}\trans \hat{\bg}_{mk}^\ast|^2/\norm{\hat{\bg}_{mk}}^4} \nonumber \\
&\quad= 1 + \EX{\frac{1}{\norm{\hat{\bg}_{mk}}^4}\hat{\bg}\trans_{mk}\EX{\tilde{\bg}^\ast_{mk}\tilde{\bg}\trans_{mk}\Big\rvert \hat{\bg}_{mk}}\hat{\bg}^\ast_{mk}} \nonumber \\
&\quad= 1+\frac{1}{N-1} \left(\frac{\beta_{mk}}{\gamma_{mk}}-1\right),
\end{align}
where in the last step we have used that $\EX{{1}/{\norm{\hat{\bg}_{mk}}^2}} = {1}/{((N-1) \gamma_{mk})}.$
Consider two different users identified by the indices $k$ and $j$, $j \neq k$. It holds that,
\begin{align}
\EX{\!\frac{ |\bg_{mk}\trans \hat{\bg}_{mj}^\ast|^2}{\norm{\hat{\bg}_{mj}}^4}\!}\! &=\! 
\begin{cases}\!\!
\dfrac{\beta_{mk}^2}{\beta_{mj}^2}\EX{\dfrac{ |\bg_{mk}\trans \hat{\bg}_{mk}^\ast|^2}{\norm{\hat{\bg}_{mk}}^4}}, &\; \text{if } \bvphi_k \!=\! \bvphi_j,  \\
\dfrac{\beta_{mk}}{(N\!-\!1)\gamma_{mj}}, &\; \text{otherwise,}
\end{cases} \nonumber \\
\!&=\! 
\begin{cases}\!\!
\dfrac{\beta_{mk}^2}{\beta_{mj}^2}\dfrac{N\!-\!2}{N\!-\!1}\!+\! \dfrac{\beta_{mk}}{(N\!-\!1)\gamma_{mj}}, &\, \text{if } \bvphi_k \!=\! \bvphi_j,  \\
\dfrac{\beta_{mk}}{(N\!-\!1)\gamma_{mj}}, &\, \text{otherwise,}
\end{cases} \label{eq:expectation} \\
\EX{\frac{ \bg_{mk}\trans \hat{\bg}_{mj}^\ast}{\norm{\hat{\bg}_{mj}}^2}} \!&= \!
\begin{cases}
\dfrac{\beta_{mk}}{\beta_{mj}}, &\quad \text{if } \bvphi_k = \bvphi_j,  \\
0, &\quad \text{otherwise.}
\end{cases}
\end{align}
In these equalities, if $\bvphi_k \! \neq \! \bvphi_j$, then we exploit that $\bg_{mk}$ is independent of $\hat{\bg}_{mj}$, else if $\bvphi_k \! = \! \bvphi_j$ we exploit the relationships among contaminated channel estimates and their mean-squares in~\eqref{eq:correlated-estimates} and~\eqref{eq:correlated-gammas}, respectively.
By using the results above, we can compute in closed form
\begin{align} 
&\mathsf{DS}_k \nonumber \\
&\quad= \sqrt{\Pd}~\sum^M_{m=1} \sqrt{\eta_{mk}}~\EX{\frac{ \bg\trans_{mk} \hat{\bg}_{mk}^\ast}{\norm{\hat{\bg}_{mk}}^2}} = \sqrt{\Pd}~\sum\limits^M_{m=1} \sqrt{\eta_{mk}}, \label{eq:ECB:DS} \\
&\EX{\left| \mathsf{BU}_k\right|^2} \nonumber \\ 
&\quad= \Pd \sum_{m=1}^M \eta_{mk}~\left( \EX{\frac{|\bg\trans_{mk} \hat{\bg}_{mk}^\ast|^2}{\norm{\hat{\bg}_{mk}}^4}} - \left| \EX{\frac{ \bg\trans_{mk} \hat{\bg}_{mk}^\ast}{\norm{\hat{\bg}_{mk}}^2}} \right|^2 \right) \nonumber \\
&\quad= \dfrac{\Pd}{N-1} \sum_{m=1}^M \eta_{mk} \left(\dfrac{\beta_{mk}}{\gamma_{mk}}-1\right), \label{eq:ECB:BU} \\
&\EX{|\mathsf{UI}_{kj}|^2} \nonumber \\
&\quad = \Pd \sum\limits_{m=1}^M \eta_{mj} \EX{\frac{|\bg\trans_{mk} \hat{\bg}_{mj}^\ast|^2}{\norm{\hat{\bg}_{mj}}^4}} \nonumber \\ 
&\qquad+ \Pd \sum\limits_{m=1}^M \sum\limits_{n\neq m}^M \sqrt{\eta_{mj} \eta_{nj}}~\EX{\frac{\bg\trans_{mk} \hat{\bg}_{mj}^\ast}{\norm{\hat{\bg}_{mj}}^2}\frac{(\bg\trans_{nk} \hat{\bg}_{nj}^\ast)^\ast}{\norm{\hat{\bg}_{nj}}^2}} \nonumber \\
&\quad= \frac{\Pd}{N\!-\!1} \sum\limits_{m=1}^M \eta_{mj} \frac{\beta_{mk}}{\gamma_{mj}}\!+\!\Pd \frac{N\!-\!2}{N\!-\!1} \sum\limits_{m=1}^M \eta_{mj}\frac{\beta^2_{mk}}{\beta^2_{mj}}~|\bvphi\herm_k \bvphi_j|^2 \nonumber \\
&\qquad + \Pd \sum\limits_{m=1}^M \sum\limits_{n\neq m}^M \sqrt{\eta_{mj} \eta_{nj}}~\frac{\beta_{mk}\beta_{nk}}{\beta_{mj}\beta_{nj}}~|\bvphi\herm_k \bvphi_j|^2, \label{eq:ECB:UI}  
\end{align}
where in the last equality we exploit the independence of channel responses and channel estimates of different APs ($n \neq m$), and the fact that the term $$\frac{\Pd}{N-1} \sum\limits_{m=1}^M \eta_{mj} \frac{\beta_{mk}}{\gamma_{mj}}$$ appears in both cases whether $\bvphi_k = \bvphi_j$ or not, as shown in~\eqref{eq:expectation}. Hence, this term does not depend on $|\bvphi\herm_k \bvphi_j|^2$. By inserting the results in~\eqref{eq:ECB:DS}--\eqref{eq:ECB:UI} into~\eqref{eq:SINR} we obtain~\eqref{eq:SINR:ECB}.

\subsection{Proof of~\eqref{eq:SE:dlpilots}}
In this section, we include a proof for the closed-form expression of the achievable downlink rate in~\eqref{eq:SE:dlpilots}. This consists in showing how both~\eqref{eq:kappa} and~\eqref{eq:SINR:CBDL} are obtained. The MMSE downlink channel estimate is given by~\cite{Interdonato2019b}
\begin{equation} \label{eq:akk_estimate}
\hat{a}_{kk}=\EX{a_{kk}} + \frac{\cov{a_{kk},\check{y}_{\mathrm{dp},k}}}{\varx{\check{y}_{\mathrm{dp},k}}}(\check{y}_{\mathrm{dp},k}-\EX{\check{y}_{\mathrm{dp},k}}),
\end{equation}
and its variance is
\begin{equation} \label{eq:var:akk_estimaition}
\kappa_k = \varx{\hat{a}_{kk}} = \frac{|\cov{a_{kk},\check{y}_{\mathrm{dp},k}}|^2}{\varx{\check{y}_{\mathrm{dp},k}}}.
\end{equation}
By following the same approach as in~\cite{Interdonato2019b}, we have
\begin{align}
%\EX{a_{kk}} &= N \sum\nolimits_{m=1}^M \sqrt{\eta_{mk}} \gamma_{mk}, \label{eq:mean:akk} \\
&\cov{a_{kk},\check{y}_{\mathrm{dp},k}} = N \sqrt{\taudp \Pdp} \sum\limits_{m=1}^M \eta_{mk} \beta_{mk} \gamma_{mk}, \label{eq:cov:akk_yk}  \\
&\varx{\check{y}_{\mathrm{dp},k}}\! = \!N \taudp \Pdp \sum\limits_{j=1}^K \sum\limits_{m=1}^M \eta_{mj} \beta_{mk} \gamma_{mj} |\bpsi_k\herm \bpsi_j|^2 \!+\! 1, \label{eq:var:yk}
%\EX{\check{y}_{\mathrm{dp},k}} &= N \sqrt{\taudp \Pdp} \sum\nolimits_{m=1}^M \sqrt{\eta_{mk}} \gamma_{mk}, \label{eq:mean:yk}
\end{align}
where equations~\eqref{eq:cov:akk_yk},~\eqref{eq:var:yk} hold when imposing~\eqref{eq:DLpilot-assignment}. By inserting equations~\eqref{eq:cov:akk_yk},~\eqref{eq:var:yk} into~\eqref{eq:var:akk_estimaition}, we obtain~\eqref{eq:kappa}. 
The mean-square of the downlink channel estimate is given by
\begin{align} \label{eq:meansq:akk_estimation}
\EX{|\hat{a}_{kk}|^2} &= \varx{\hat{a}_{kk}} + |\EX{\hat{a}_{kk}} |^2 \nonumber \\
&= \kappa_k + \left(N \sum\limits_{m=1}^M \sqrt{\eta_{mk}} \gamma_{mk} \right)^2,
\end{align}
where $$ \EX{\hat{a}_{kk}} = \EX{a_{kk}} = N \sum\limits_{m=1}^M \sqrt{\eta_{mk}} \gamma_{mk}, $$ as the MMSE estimator is unbiased under the \textit{regularity assumptions}.
The mean-square of the downlink channel estimation error is given by
\begin{align} \label{eq:meansq:akk_error_estimation}
\EX{|\tilde{a}_{kk}|^2} &= \EX{|a_{kk}-\hat{a}_{kk}|^2} \nonumber \\
&= \EX{|{a}_{kk}|^2} + \EX{|\hat{a}_{kk}|^2} - 2\operatorname{Re}(\EX{a^{\ast}_{kk}\hat{a}_{kk}}) \nonumber \\
&\stackrel{(a)}{=} \EX{|{a}_{kk}|^2} - \EX{|\hat{a}_{kk}|^2} \stackrel{(b)}{=} N\varsigma_{kk}-\kappa_k,
\end{align}
where $(a)$ results from
\begin{align}
\EX{a^{\ast}_{kk}\hat{a}_{kk}} &= |\EX{a_{kk}}|^2 + \frac{\cov{a_{kk},\check{y}_{\mathrm{dp},k}}}{\varx{\check{y}_{\mathrm{dp},k}}} \nonumber \\ &\qquad\times(\EX{a^{\ast}_{kk}\check{y}_{\mathrm{dp},k}}-\EX{a_{kk}}^{\ast}\EX{\check{y}_{\mathrm{dp},k}}) \nonumber \\
&= |\EX{a_{kk}}|^2 + \frac{|\cov{a_{kk},\check{y}_{\mathrm{dp},k}}|^2}{\varx{\check{y}_{\mathrm{dp},k}}} \nonumber \\&= \EX{|\hat{a}_{kk}|^2},
\end{align} 
while $(b)$ follows from~\eqref{eq:meansq:akk_estimation} and the fact that 
\begin{align} \label{eq:meansq:akk}
\!\!\EX{|{a}_{kk}|^2}\!\! =\! N \!\sum\limits_{m=1}^M \eta_{mk} \beta_{mk} \gamma_{mk} \! +\! \left(\!N\! \sum\limits_{m=1}^M \!\!\sqrt{\eta_{mk}} \gamma_{mk}\! \right)^2\!\!.
\end{align}
Finally, we focus on deriving $\EX{|{a}_{kj}|^2}$, $j \neq k$, in closed form. Under the same assumptions on the channel vectors considered in Appendix \textit{A}, it holds that
\begin{align}
&\EX{|\bg_{mk}\trans \hat{\bg}_{mk}^\ast|^2} \nonumber \\
&\quad= \EX{|(\hat{\bg}_{mk} + \tilde{\bg}_{mk})\trans \hat{\bg}_{mk}^\ast|^2} \nonumber \\
&\quad= \EX{|\hat{\bg}_{mk}\trans \hat{\bg}_{mk}^\ast|^2} +  \EX{\hat{\bg}\trans_{mk}\EX{\tilde{\bg}^\ast_{mk}\tilde{\bg}\trans_{mk}\Big\rvert \hat{\bg}_{mk}}\hat{\bg}^\ast_{mk}} \nonumber \\
&\quad= (N+1)N \gamma_{mk}^2 + N (\beta_{mk}-\gamma_{mk})\gamma_{mk} \nonumber \\
&\quad= N^2 \gamma_{mk}^2 + N \beta_{mk} \gamma_{mk}, \\
&\EX{|\bg\trans_{mk} \hat{\bg}^{\ast}_{mj}|^2} \nonumber \\
&\quad= 
\begin{cases}
\dfrac{\beta_{mj}^2}{\beta_{mk}^2} \EX{|\bg_{mk}\trans \hat{\bg}_{mk}^\ast|^2}, &\quad \text{if } \bvphi_k \!=\! \bvphi_j,  \\
N \beta_{mk} \gamma_{mj}, &\quad \text{otherwise}
\end{cases} \nonumber \\
&\quad=
\begin{cases}
N^2 \gamma_{mk}^2 \dfrac{\beta_{mj}^2}{\beta_{mk}^2} + N \beta_{mk} \gamma_{mj}, &\quad \text{if } \bvphi_k \!=\! \bvphi_j,  \\
N \beta_{mk} \gamma_{mj}, &\quad \text{otherwise}
\end{cases} \nonumber \\
&\quad= N^2 \gamma_{mj}^2 \dfrac{\beta_{mk}^2}{\beta_{mj}^2} |\bvphi\herm_k\bvphi_j|^2 + N \beta_{mk} \gamma_{mj},  \label{eq:meansq:inner-product} \\
&\EX{\bg\trans_{mk} \hat{\bg}^{\ast}_{mj}(\bg\trans_{nk} \hat{\bg}^{\ast}_{nj})^{\ast}} \nonumber \\
&\quad= \EX{\bg\trans_{mk} \hat{\bg}^{\ast}_{mj}}\EX{\bg\trans_{nk} \hat{\bg}^{\ast}_{nj}} \nonumber \\ 
&\quad= 
\begin{cases}
N^2 \dfrac{\beta_{mk}}{\beta_{mj}} \dfrac{\beta_{nk}}{\beta_{nj}} \gamma_{mj} \gamma_{nj}, &\quad \text{if } \bvphi_k \!=\! \bvphi_j,  \\
0, &\quad \text{otherwise}
\end{cases} \nonumber \\
&\quad= N^2 \dfrac{\beta_{mk}}{\beta_{mj}} \dfrac{\beta_{nk}}{\beta_{nj}} \gamma_{mj} \gamma_{nj} |\bvphi\herm_k\bvphi_j|^2. \label{eq:mean:cross-products}
\end{align}
In these equalities, if $\bvphi_k \! \neq \! \bvphi_j$, then we exploit that $\bg_{mk}$ is independent of $\hat{\bg}_{mj}$, else if $\bvphi_k \! = \! \bvphi_j$ we exploit the relationships among contaminated channel estimates and their mean-squares in~\eqref{eq:correlated-estimates} and~\eqref{eq:correlated-gammas}, respectively. Moreover, in~\eqref{eq:mean:cross-products}, we exploit the independence of channel responses and channel estimates of different APs ($n \neq m$). 
By using the results in equations~\eqref{eq:meansq:inner-product},~\eqref{eq:mean:cross-products}, we can compute in closed form
\begin{align} \label{eq:meansq:akj}
\EX{|{a}_{kj}|^2} &= \sum\limits_{m=1}^M \eta_{mj} \EX{|\bg\trans_{mk} \hat{\bg}^{\ast}_{mj}|^2} \nonumber \\ 
&\quad+ \sum\limits_{m=1}^M \sum\limits_{n \neq m}^M \sqrt{\eta_{mj}\eta_{nj}} \EX{\bg\trans_{mk} \hat{\bg}^{\ast}_{mj}(\bg\trans_{nk} \hat{\bg}^{\ast}_{nj})^{\ast}} \nonumber \\
&= N \sum\limits_{m=1}^M \eta_{mj} \beta_{mk} \gamma_{mj} \nonumber \\
&\quad  + \left(N \sum\limits_{m=1}^M \sqrt{\eta_{mj}} \gamma_{mj} \frac{\beta_{mk}}{\beta_{mj}} \right)^2 |\bvphi\herm_k\bvphi_j|^2.  
\end{align}
By inserting~\eqref{eq:meansq:akk_estimation},~\eqref{eq:meansq:akk_error_estimation} and~\eqref{eq:meansq:akj} into~\eqref{eq:SINR:CBDL:general}, we finally obtain~\eqref{eq:SINR:CBDL} and in turn~\eqref{eq:SE:dlpilots}.

\bibliographystyle{IEEEtran}
\bibliography{IEEEabrv,refs-abbr} 

\vspace*{-10mm}
\begin{IEEEbiography}[{\includegraphics[width=1in,height=1.25in,clip,keepaspectratio]{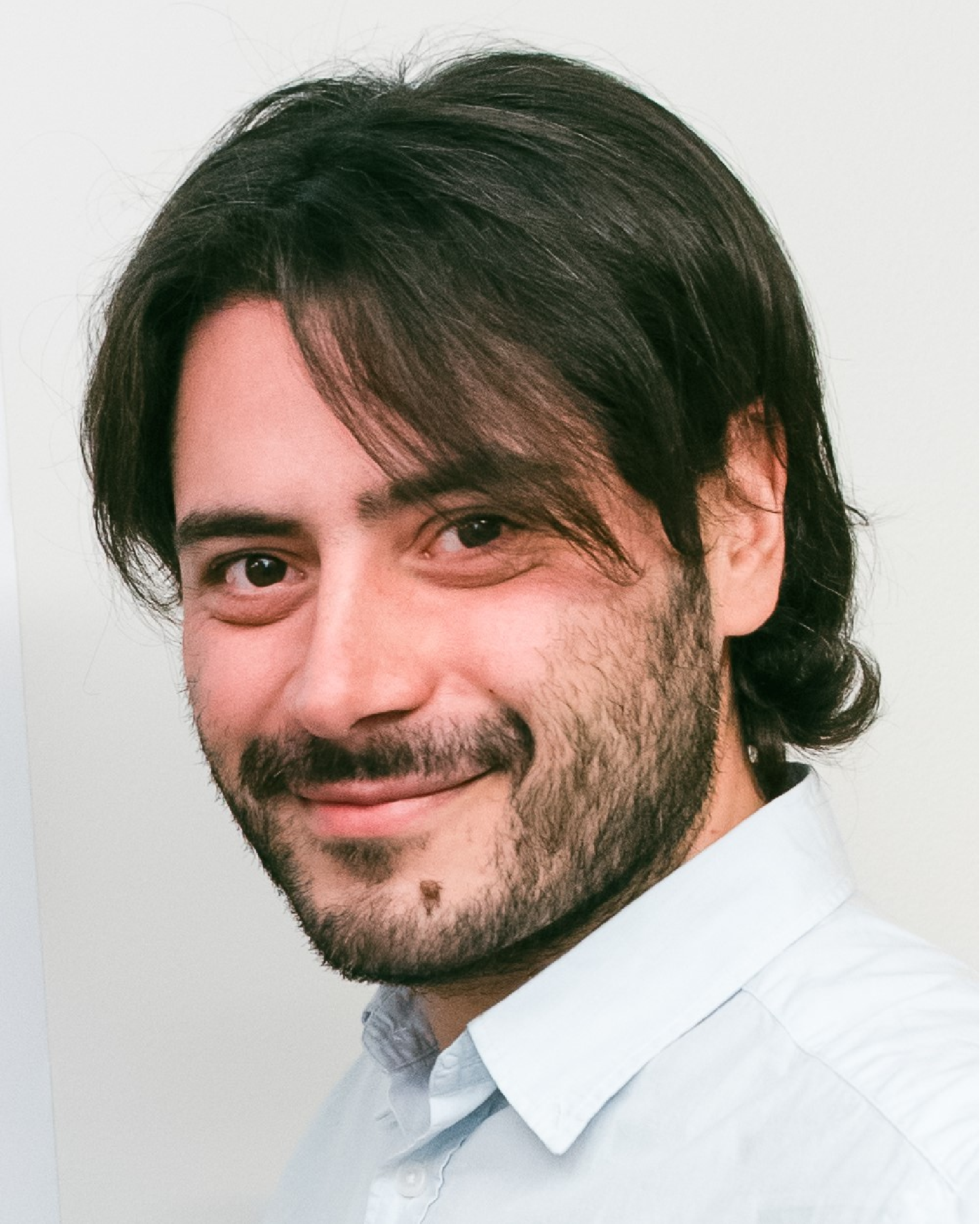}}]{Giovanni Interdonato}
(Member, IEEE) received the M.Sc. degree in computer and telecommunication systems engineering from the University Mediterannea of Reggio Calabria, Italy, in 2015, and the Ph.D. degree in electrical engineering with specialization in communication systems from Link\"oping University (LiU), Sweden, in 2020. From October 2015 to October 2018, he was researcher at the radio network department at Ericsson Research in Link\"oping, and a Marie Sklodowska-Curie research fellow of the H2020 ITN project \textit{5Gwireless}. 
He is currently a postdoctoral researcher at the Department of Electrical and Information Engineering, University of Cassino and Southern Lazio, Italy. 
His main research interests include distributed (cell-free) Massive MIMO systems, and 5G New Radio communication protocols. 

He has filed about twenty Massive MIMO related patent applications, and received a scholarship from the Ericsson Research Foundation in 2019.     
\end{IEEEbiography}

\begin{IEEEbiography}[{\includegraphics[width=1in,height=1.25in,clip,keepaspectratio]{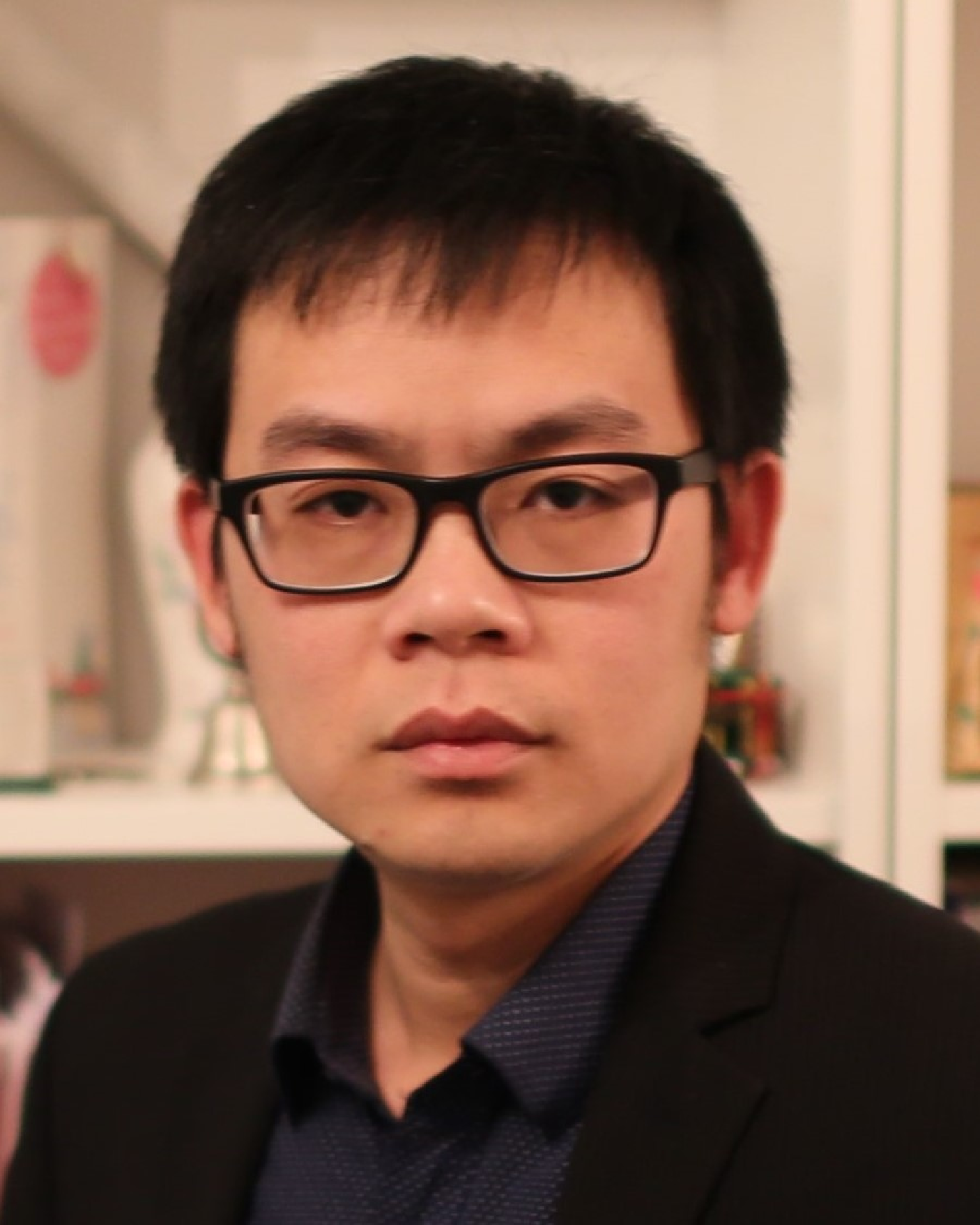}}]{Hien Quoc Ngo} 
(Senior Member, IEEE) received the B.S. degree in electrical engineering from the Ho Chi Minh City University of Technology, Vietnam, in 2007, the M.S. degree in electronics and radio engineering from Kyung Hee University, South Korea, in 2010, and the Ph.D. degree in communication systems from Link\"oping University (LiU), Sweden, in 2015. In 2014, he visited the Nokia Bell Labs, Murray Hill, New Jersey, USA. From January 2016 to April 2017, Hien Quoc Ngo was a VR researcher at the Department of Electrical Engineering (ISY), LiU. He was also a Visiting Research Fellow at the School of Electronics, Electrical Engineering and Computer Science, Queen's University Belfast, UK, funded by the Swedish Research Council.

Hien Quoc Ngo is currently a Lecturer at Queen's University Belfast, UK. His main research interests include massive (large-scale) MIMO systems, cell-free massive MIMO, physical layer security, and cooperative communications. He has co-authored many research papers in wireless communications and co-authored the Cambridge University Press textbook \emph{Fundamentals of Massive MIMO} (2016).

Dr. Hien Quoc Ngo received the IEEE ComSoc Stephen O. Rice Prize in Communications Theory in 2015, the IEEE ComSoc Leonard G. Abraham Prize in 2017, and the Best PhD Award from EURASIP in 2018. He also received the IEEE Sweden VT-COM-IT Joint Chapter Best Student Journal Paper Award in 2015. He was an \emph{IEEE Communications Letters} exemplary reviewer for 2014, an \emph{IEEE Transactions on Communications} exemplary reviewer for 2015, and an \emph{IEEE Wireless Communications Letters} exemplary reviewer for 2016.  He was awarded the UKRI Future Leaders Fellowship in 2019.
Dr. Hien Quoc Ngo currently serves as an Editor for the IEEE Transactions on Wireless Communications, IEEE Wireless Communications Letters, Digital Signal Processing, Elsevier Physical Communication (PHYCOM), and IEICE Transactions on Fundamentals of Electronics, Communications and Computer Sciences. He was a Guest Editor of IET Communications, special issue on ``Recent Advances on 5G Communications'' and a Guest Editor of  IEEE Access, special issue on ``Modelling, Analysis, and Design of 5G Ultra-Dense Networks'', in 2017. He has been a member of Technical Program Committees for several IEEE conferences such as ICC, GLOBECOM, WCNC, and VTC.
\end{IEEEbiography}

\enlargethispage{-4in}
\begin{IEEEbiography}[{\includegraphics[width=1in,height=1.25in,clip,keepaspectratio]{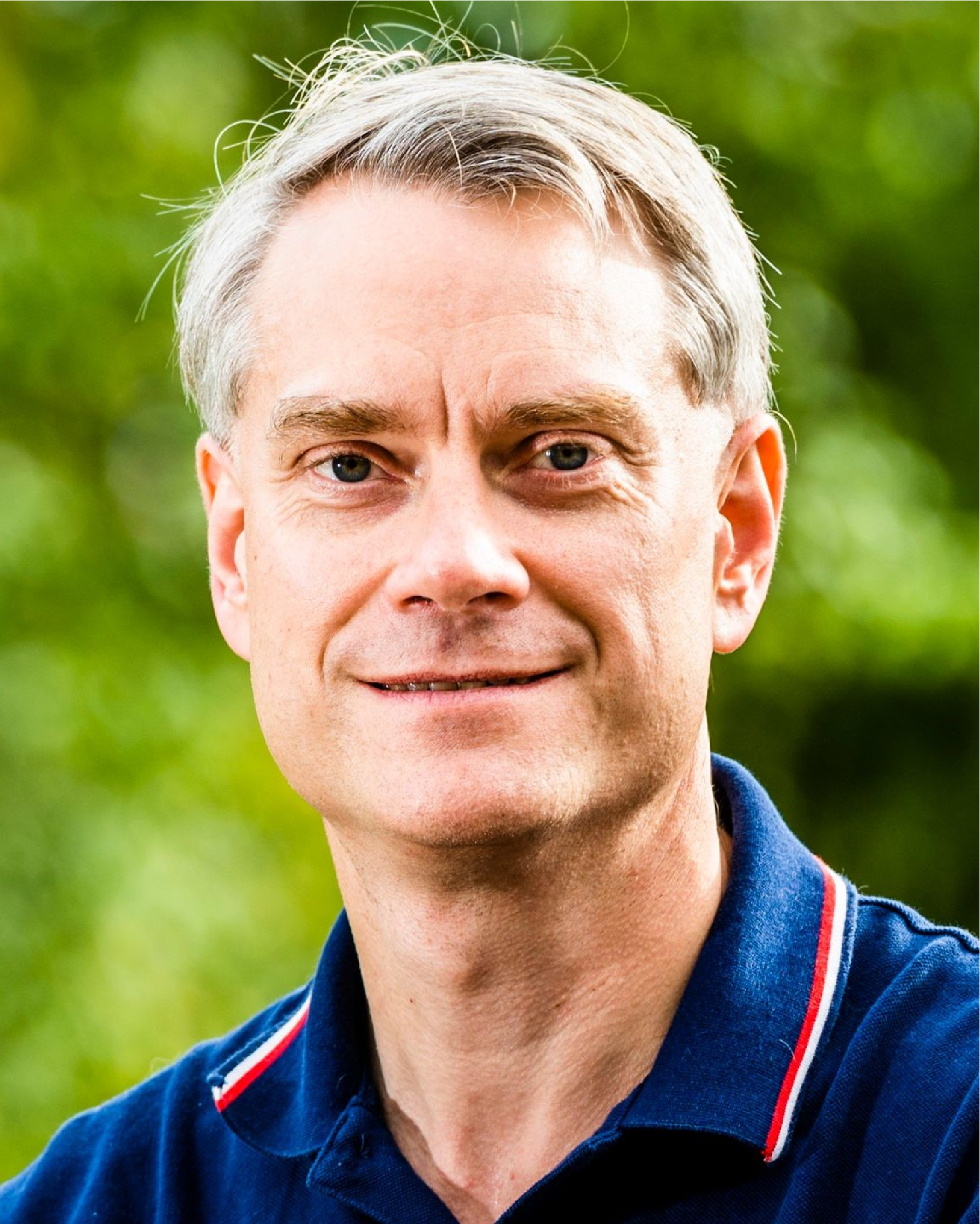}}]{Erik G. Larsson}
(Fellow, IEEE) received the Ph.D. degree from Uppsala University,
Uppsala, Sweden, in 2002.  He is currently Professor of Communication
Systems at Link\"oping University (LiU) in Link\"oping, Sweden. He was
with the KTH Royal Institute of Technology in Stockholm, Sweden, the
George Washington University, USA, the University of Florida, USA, and
Ericsson Research, Sweden.  His main professional interests are within
the areas of wireless communications and signal processing. He 
co-authored \emph{Space-Time Block Coding for  Wireless Communications} (Cambridge University Press, 2003) 
and \emph{Fundamentals of Massive MIMO} (Cambridge University Press, 2016). 

Currently he is a member of the  \emph{IEEE Transactions on Wireless Communications}    steering committee. 
He served as  chair  of the IEEE Signal Processing Society SPCOM technical committee (2015--2016), 
  chair of  the \emph{IEEE Wireless  Communications Letters} steering committee (2014--2015), 
    General and Technical Chair of the Asilomar SSC conference (2015, 2012), 
  technical co-chair of the IEEE Communication Theory Workshop (2019), 
  and   member of the  IEEE Signal Processing Society Awards Board (2017--2019).
He was Associate Editor for, among others, the \emph{IEEE Transactions on Communications} (2010-2014), 
 the \emph{IEEE Transactions on Signal Processing} (2006-2010),
and  the \emph{IEEE Signal
  Processing Magazine} (2018-2020).
  
He received the IEEE Signal Processing Magazine Best Column Award
twice, in 2012 and 2014, the IEEE ComSoc Stephen O. Rice Prize in
Communications Theory in 2015, the IEEE ComSoc Leonard G. Abraham
Prize in 2017, the IEEE ComSoc Best Tutorial Paper Award in 2018, and
the IEEE ComSoc Fred W. Ellersick Prize in 2019.
\end{IEEEbiography}

\end{document}